\documentclass[12pt,preprint]{aastex}

\slugcomment{Accepted for Publication in The Astrophysical Journal}

\shorttitle{Damped Ly-$\alpha$ Absorbers}
\shortauthors{Kulkarni et al.}

\begin{document}


\title{Hubble Space Telescope Observations of Element 
Abundances in Low-redshift Damped Lyman-$\alpha$ Galaxies and 
Implications for the Global Metallicity-Redshift Relation  
}


\author{Varsha P. Kulkarni}
\affil{Dept. of Physics \& Astronomy, University of South Carolina,  
Columbia, SC 29208}

\author{S. Michael Fall}
\affil{Space Telescope Science Institute, Baltimore, 
MD 21218}

\author{James T. Lauroesch}
\affil{Department of Physics \& Astronomy, Northwestern University, Evanston, IL 60208}

\author{Donald G. York \altaffilmark{1} and Daniel E. Welty} 
\affil{Dept. of Astronomy \& Astrophysics, University of Chicago, Chicago, IL 60637}

\author{Pushpa Khare}
\affil{Dept. of Physics, Utkal University, Bhubaneswar, India}

\author{James W. Truran \altaffilmark{1}} 
\affil{Dept. of Astronomy \& Astrophysics, University of Chicago, Chicago, IL 60637}

\altaffiltext{1}{Also, the Enrico Fermi Institute, University of Chicago, Chicago, IL 60637}



\begin{abstract}
Most models of cosmic chemical evolution predict that 
the mass-weighted mean interstellar 
metallicity of galaxies should rise with time from a low value $\sim 1/30$ 
solar at $z \sim 3$ to a nearly solar value at $z = 0$. In the absence of 
any selection effects, the damped Lyman-alpha absorbers (DLAs) in quasar 
spectra 
are expected to show such a rise in global metallicity. However, it 
has been difficult to determine whether or not DLAs show this effect, 
primarily because of the very small number of DLA metallicity measurements 
at low redshifts. 

In an attempt to put tighter constraints on the low-redshift end of the 
DLA metallicity-redshift relation, we have observed  
Zn II and Cr II lines in four DLAs at $0.09 < z < 0.52$, using the Space Telescope 
Imaging Spectrograph (STIS) onboard the Hubble Space Telescope (HST). 
These observations have provided the first constraints on Zn abundances 
in DLAs with $z < 0.4$. 
In all the three DLAs for which our observations offer meaningful constraints on 
the metallicity, the data suggest that the metallicities are much lower 
than the solar value. 
These results are consistent with recent imaging studies indicating that  
these DLAs may be associated with dwarf or low surface brightness galaxies. 

We combine our results with higher redshift data from the literature   
to estimate the global mean metallicity-redshift relation for DLAs. 
We find that the global mean metallicity shows at most a slow increase 
with decreasing redshift. For the redshift range $0.09 < z < 3.90$, 
the slope of the exponential fit to the binned 
$N({\rm H \, I})$-weighted  mean Zn metallicity 
vs. redshift relation is $ -0.18 \pm 0.06$ counting Zn limits as 
detections, $-0.22 \pm 0.08$ counting Zn limits as zeros, and 
$ -0.23 \pm 0.06$ using constraints on metallicity from 
other elements in cases of Zn limits. 
The corresponding estimates of the $z = 0$ intercept of the metallicity-redshift 
relation are $-0.74 \pm 0.15$, $-0.75 \pm 0.18$, and $-0.71 \pm 0.13$, respectively. 
Roughly similar results are obained if survival analysis or an unbinned 
$N({\rm H \, I})$-weighted nonlinear $\chi^{2}$ approach is used. Thus,  
the $N({\rm H \, I})$-weighted mean metallicity of DLAs does not appear 
to rise up to solar or near-solar values at $z =0$. This weak  
evolution could be explained by the fact that our 
absorption-selected sample seems to be dominated by dwarf or low surface brightness 
galaxies. This suggests that current DLA samples, especially those 
at low redshifts, could be biased against more  
enriched galaxies because the latter may cause higher dust obscuration of the 
background quasars. 
\end{abstract}





\section{Introduction}

A great deal of progress has been made in recent years in the observations 
of high-redshift galaxies. The average star formation history of the 
Universe has been estimated from emission properties 
of galaxies detected in deep imaging and redshift surveys (e.g., Lilly et al. 
1996; 
Madau et al. 1996). This emission history of galaxies is tied 
intimately to the history of gas consumption and metal production 
in galaxies, because the global mean densities of gas, metals, 
and stars are coupled through conservation-type 
relations. Constraining the history of metal production thus 
puts constraints on the histories of star formation and gas consumption 
(e.g., Pei \& Fall 1995).

Most models of cosmic chemical evolution, ranging from analytical 
calculations to numerical simulations, predict the mass-weighted 
mean global metallicity 
in galaxies, given by 
$\Omega_{\rm{metals}}^{\rm ISM}/\Omega_{\rm{gas}}^{\rm ISM}$,  
to rise from a low value at high redshift to a near-solar 
value at zero redshift (e.g., Pei \& Fall 1995; Malaney \& Chaboyer 1996; 
Pei, Fall, \& Hauser 1999; Somerville, Primack, \& Faber 2001; Tissera et al. 2001). 
[Here $\Omega_{\rm{metals}}^{\rm ISM}$ and $\Omega_{\rm{gas}}^{\rm ISM}$ 
refer, respectively, to the mean comoving density of metals and the total 
mean comoving density of gas in the interstellar matter of galaxies, 
both measured in units of the present critical density ($\rho_{c} = 
3 H_{0}^{2}/{8 \pi G}$).]
The present-day mass-weighted mean metallicity of local 
galaxies is also observed to be close to solar (e.g., Kulkarni \& Fall 2002). 
It is therefore of great interest to determine whether element abundance 
measurements in galaxies actually show the predicted rise in mean metallicity 
with time.

Abundance measurements in quasar absorption systems can directly probe 
the history of metal production in galaxies. 
Particularly useful in this respect are the damped Lyman-$\alpha$ (DLA) 
absorbers (log $N_{\rm H I} \ge 20.3$). Under reasonable assumptions, 
DLAs may contain enough neutral gas to 
form a large fraction of the stars visible today (Wolfe et al. 1995). 
DLAs are our principal source of information about the 
chemical content of interstellar matter in high-$z$ galaxies  
(e.g., Meyer \& York 1992; Lu et al. 1996; Kulkarni et al. 1996, 
1997, 1999; Pettini et al. 1994, 1997, 1999; Prochaska \& 
Wolfe 1999; Prochaska et al. 2001a, 2003a, 2003b; and 
references therein). 

For a variety of reasons, the abundance of Zn is often taken as the 
metallicity indicator in DLAs. Zn tracks Fe closely, to within 
$\sim \pm 0.1$ dex, in Galactic halo and disk stars 
for metallicities [Fe/H] $\ga -3$ (e.g., Mishenina et al. 2002; 
Cayrel et al. 2004; Nissen et al. 2004). Zn is also relatively 
undepleted on interstellar dust grains, so that 
gas-phase Zn abundances  can give metallicity 
estimates relatively free of depletion effects. Finally, Zn II has 
two absorption lines at $\lambda \lambda 2026, 2062$ that, in DLAs, 
often lie outside the Lyman-$\alpha$ forest and are usually unsaturated. 
There are, however, some uncertainties related to the nucleosynthesis 
of Zn and the potential role of ionization corrections, and we return 
to these issues in section 5.1.

The absorption lines of DLAs sample the interstellar matter of 
distant galaxies. This sampling is completely independent of many properties 
of the galaxies, including their luminosities, sizes, and morphologies. 
It can be shown 
(e.g., Lanzetta, Wolfe, \& Turnshek 1995; Kulkarni \& Fall 2002) that, 
in the absence of selection effects, the quantity 
$\Omega_{\rm{metals}}^{\rm ISM}/\Omega_{\rm{gas}}^{\rm ISM}$ in galaxies 
is equal to the   
$N_{\rm H I}$-weighted mean metallicity $\overline{Z}$ in a sample 
of DLAs,  given by 
\begin{equation}
\overline{Z} = 
{\Sigma N({\rm Zn \, II})_{i} / \Sigma N({\rm H \, I})_{i} \over 
{({\rm Zn/H})_{\odot}}} Z_{\odot}.   
\end{equation} 
This quantity $\overline{Z}$ may therefore be expected to show a rise from a low value at 
high redshift to a near-solar value at low redshift. 

There has been great debate about this issue in the 
past few years, with most studies advocating no evolution in the global 
mean metallicity $\overline{Z}$ inferred from DLAs (e.g., Pettini et al. 1997, 1999; Prochaska 
\& Wolfe 1999, 2000; Vladilo et al. 2000; Savaglio 2001; Prochaska, 
Gawiser, \& Wolfe 2001b). Kulkarni \& Fall (2002) examined this 
issue further 
by applying various statistical analyses to the Zn data from the literature. 
Based on a compilation of 57 Zn measurements (36 detections and 21 limits) 
from the literature at redshifts $0.4 < z < 3.4$, Kulkarni \& Fall (2002) 
found the slope of the 
metallicity-redshift relation to be $-0.26 \pm 0.10$, consistent 
at $\approx 2-3 \, \sigma$ level with both the predicted rates of evolution 
(-0.25 to -0.61), and with no evolution. Recently, Prochaska et al. (2003b) 
have reached similar conclusions (slope $-0.25 \pm 0.07$) for $0.5 < z < 4.7$, 
using 
Zn measurements in 11 DLAs and Fe, Si, S, or O measurements in 110 other DLAs.  
As in the case of Zn, the elements Fe, Si, S, and O 
also have both advantages (e.g., better theoretical understanding of nucleosynthesis, 
ease of detection) and disadvantages [e.g., dust depletion (for Fe, Si), 
lack of two or more 
weak lines outside the Lyman-alpha forest (for Si, S, O), potential 
ionization effects (for Si, S), potential enhancement with respect 
to the Fe-group elements due to alpha enrichment (for Si, S, O), etc.]. 
In any case, the results of Prochaska et al. (2003b), based on these 
other elements, are similar to those of Kulkarni \& Fall (2002), as 
discussed further in section 4. 

The issue of metallicity evolution in DLAs is still not fully 
understood. The main reason for the debate about this issue 
is the small number of measurements available, especially at $z < 1.5$. 
The samples of Kulkarni \& Fall (2002) and Prochaska et al. 
(2003b) both include only 1 measurement each in the redshift range 
$z \la 0.6$, which corresponds to $> 40 \%$ of the age of the Universe. 
The low-redshift systems require space-based ultraviolet (UV) measurements 
to access the H I lines at $z < 1.6$ and the Zn II lines at $z < 0.6$. 
With the goal of putting better constraints on the low-redshift end of the DLA 
metallicity-redshift relation,
we have obtained UV spectra of three quasars with four intervening DLAs at 
$0.09 < z < 0.52$ 
using the Space Telescope Imaging Spectrograph (STIS) onboard the 
{\it Hubble Space Telescope (HST)}. Here we report our new observations of the 
Zn II and Cr II absorption in these low-redshift DLAs and their implications for the 
metallicity-redshift relation for DLAs. We have chosen to focus on Zn 
as the primary metallicity indicator, rather than combine measurements 
of Zn with those of Si or S, primarily 
because (a) we prefer to use the more uniform Zn sample, relatively free of 
assumptions 
about dust depletion, and (b) STIS is much more sensitive at the longer 
wavelengths of the Zn II lines than at the wavelengths of the S II and Si II lines 
for our DLAs. Section 2 describes our observations 
and data reduction. Section 3 describes the results for the individual 
DLAs. Section 4 presents 
the implications of our results, combined with higher redshift data from the 
literature, for the metallicity-redshift relation of DLAs. 
Section 5 discusses the advantages and disadvantages of using Zn, the 
morphological information for the galaxies giving rise to the DLAs in our 
sample, and potential selection effects. Finally, section 6 provides a 
summary of our results.

\section{Observations and Data Reduction}

\subsection{Observations}
Our sample consists of 4 confirmed DLAs with $0.09 < z < 0.52$ and ${\rm log} 
N_{\rm H I} 
\ga 20.3$ in the sightlines toward the quasars Q0738+313, Q0827+243, 
and Q0952+179. Table 1 lists the key properties of these DLAs known from 
previous studies. 
The lines of interest are Zn II $\lambda \, \lambda$ 2026.136, 
2062.664, and Cr II $\lambda \, \lambda$ 2056.254, 2062.234, 2066.161, which 
occur in the range of 2211-3145 {\AA} for the DLAs in our sample. 

Our observations 
were obtained during February-November 2003 using {\it HST} STIS. Table 2 
summarizes the details of 
our observations. The STIS near-ultraviolet 
Multi-Anode Microchannel Array (nuv-MAMA) was used for Q0738+313 and Q0952+179. 
For Q0827+243, the CCD was used because of its higher efficiency at the 
longer wavelengths required for this object. The first-order gratings 
G230M and G230MB were used to get the moderately high spectral resolution and 
wavelength coverage necessary for the abundance determinations. 
The $52 \arcsec \times 0.2 \arcsec$ slit was used for the MAMA 
observations to achieve a good 
compromise between spectral resolution and throughput. 
For the CCD, the 52x0.2E1 pseudo-aperture (with the default target 
placement near row 893, $\sim 5 \arcsec$ from the top of the CCD) 
was used to mitigate the effects of the charge-transfer inefficiency. 

All targets were acquired with the F28x50LP aperture using the STIS CCD in 
the acquisition mode. Acquisition exposures of 3-5 s were used to achieve 
a signal-to-noise ratio (SNR) of $> 40$, taking care not to saturate the detector. The science 
exposures were dithered using the ``STIS-ALONG-SLIT'' 
pattern in 3, 4, or 5 steps of 0.4 $\arcsec$ each, so as to help with 
flatfielding and removal of hot pixels. For the CCD observations, the 
exposures at each dithered setting were divided into 3 equal subexposures  
with CR-SPLIT=3 to facilitate the rejection of cosmic rays. An automatic 
wavelength calibration exposure at the respective grating setting was included 
in each orbit. 

\subsection{Data Reduction}

All data reduction tasks were performed independently by two 
members of our team, using multiple techniques as discussed below.
The spectra were reduced using the Image Reduction and Analysis 
Facility (IRAF) version 2.12.1 and the CALSTIS package in the 
Space Telescope Science Data Analysis System (STSDAS) software version 
3.1. The automatic lamp exposures obtained during our observations were 
used for wavelength calibration.  For all other calibrations, 
the most recent reference files available in the {\it HST} data 
archives were used. These include image reference files for 
darks, pixel-to-pixel flat fields, low-order flat fields, tables 
for aperture description, aperture throughput, bad pixels, dispersion 
coefficients, incidence angle correction, and image distortion correction 
for both the MAMA and the CCD. In addition, the CCD reductions required  
the bias calibration files from the {\it HST} archives. For the MAMA, other 
calibration files included 
tables for linearity, offset correction, scattered light correction, 
near-UV dark correction, and time-dependent sensitivity correction. 

After the initial reduction of each 2-dimensional image, the spectral 
extraction was carried out using the STSDAS task X1D. 
In order to obtain optimum signal-to-noise ratios in the 1-dimensional 
spectra, a variety of extraction heights were tested.  Extraction 
heights of 3 or 4 pixels were found to encompass all of the 
recoverable flux without introducing additional noise from 
the background. The results of X1D were 
checked by repeating the spectral extractions with the IRAF task APALL.  
In these latter reductions, the sky background level was fitted in the regions 
on either side of the 
quasar spectrum, and this fit used to subtract the background level. 
In most cases, second-order Chebyshev polynomials were found to provide adequate 
fits to the traces of the spectra. The results of these various reductions 
were found to be in very close agreement. Finally, for each target, the individual 
1-dimensional spectra at each wavelength setting were combined 
together. The combined spectra were then 
continuum-fitted using Chebyshev or cubic spline functions of orders 2 to 6, depending on the shape 
of the continuum and the presence or absence of broad emission lines in the 
wavelength range observed. Figs. 1-5 show the continuum-normalized 
intensity versus vacuum wavelength for all the targets. The spectral 
dispersions 
are 0.09 {\AA} pixel$^{-1}$ for the NUV-MAMA spectra, and 0.15 {\AA} 
pixel$^{-1}$ for the CCD spectra. The resulting spectral resolutions are 
in the range of $\approx 21-24$ km s$^{-1}$ for the NUV-MAMA data and 
$\approx 29$ km s$^{-1}$ for the CCD data. 

Table 3 lists the features measured in the final spectra. The central 
wavelengths and equivalent widths were measured using 
the IRAF task SPLOT. The 1 $\sigma$ uncertainties in equivalent widths 
due to photon noise were estimated assuming Poisson statistics 
as $\sigma_{W}^{p} = \Delta \lambda \sqrt{N_{pix}}/SNR$ 
where $\Delta \lambda$ is the spectral dispersion, $N_{pix}$ is the number 
of pixels occupied by the feature, and $SNR$ is the signal-to-noise ratio 
per pixel in the continuum near the feature.  

Another uncertainty often not considered in spectroscopic studies of QSO absorption
line systems is the uncertainty in the equivalent width (and hence column
density) associated with the continuum placement (e.g., Savage \& Sembach
1991).  While the continua of QSOs are in general much simpler than the stellar
continua encountered in studies of the Galactic interstellar matter, there is still some uncertainty associated with the exact placement
of the continuum. In order to assess this uncertainty, we slid the
continuum up and down by 0.3 times the 1 $\sigma$ noise level and measured the 
equivalent widths for these extrema (see Sembach \& Savage 1992 for a 
justification of the scale factor 0.3.). 
Table 4 lists the continuum placement uncertainties in the equivalent widths 
$\sigma_{W}^{c}$ thus calculated for the DLA metal features. 
Also listed are the total uncertainties $\sigma_{W}^{t}$ calculated 
by adding the photon noise uncertainties and continuum uncertainties in quadrature. 
Further details on the line measurements are provided in section 3.

\subsection{Estimation of Column Densities}

Column densities for Zn II and Cr II were estimated by fitting
the observed absorption line profiles using the program FITS6P 
(Welty, Hobbs, \& York 1991) that evolved from the code used by  
Vidal-Madjar et al. (1977). FITS6P minimizes 
the $\chi^{2}$ between the data and the theoretical Voigt profiles 
convolved with the instrumental profile. 
The atomic data were adopted from Bergeson \& Lawler (1993) 
for Zn II $\lambda \, \lambda 
2026, 2062$ and Cr II $\lambda \, \lambda 2056, 2062, 2066$, and 
from Bergeson, Mullman, \& Lawler (1994) for 
Fe II $\lambda \, \lambda 2249, 2260$. 
The oscillator strengths used for these lines were 
0.489, 0.256 for Zn II $\lambda \, \lambda 2026, 2062$; 
0.105, 0.078, 0.0515 for Cr II $\lambda \, \lambda 2056, 2062, 2066$;  
and 0.00182, 0.00244 for Fe II $\lambda$ 2249, 2260. 

In an attempt to constrain the column densities and effective 
Doppler $b$ 
parameters, we performed simultaneous single-component fits to 
the detected lines of Zn II, Cr II, and Fe II (where available), 
varying the column densities $N$, the Doppler parameter 
$b$, and the central velocity $v$ together, but 
assuming a common 
$b$ and $v$ value for Zn II, Cr II, and the weak Fe II lines. We 
also performed simultaneous single-component fits to both detected 
and undetected lines. The 
uncertainties in $N$, $b$, and $v$ were determined simultaneously in these 
fits. 

In the cases where a 2026 {\AA} feature 
was potentially present, we investigated its contribution by 
carrying out profile fits with the Mg I $\lambda$ 2026 line included. 
To do this, the Mg I column density was   
estimated from profile fitting of the Mg I $\lambda$ 2853 feature in 
data obtained at the Multiple Mirror Telescope (MMT), if available 
(Khare et al. 2004). The MMT data were obtained as part of a larger 
survey of element abundances in DLAs at $z < 1.5$. These data have lower 
spectral resolution    
($\sim 75-95$ km s$^{-1}$) but higher SNR ($\sim 50-90$). 
The Mg I $\lambda 2026$ line was 
found to make insignificant contribution to the observed 2026 {\AA} feature. 
Indeed, there is no evidence of absorption at the expected positions of 
Mg I $\lambda$ 2026 in any of the 3 DLAs where the relevant wavelength 
region was covered. 
 
The best-fit effective $b$ values were found to be in 
the range of 11-15  km s$^{-1}$. However, due to the moderate resolution 
and SNR of our data and the 
weakness of the lines, it was difficult to derive accurate estimates 
for the effective $b$ values. Therefore, in order to explore the dependence 
of the estimated column densities on the $b$ values, we also performed 
single-component fits to the individual detected lines for several (fixed) 
choices of $b$, i.e., 5, 10, 15, and 20 km s$^{-1}$. In general, relatively 
little difference in column density was found for 
the fits with $b$-values of 5-20 km~s$^{-1}$, indicating that the lines 
would be near the linear part of 
the curve of growth for that range in effective $b$. 
To search for any broad, shallow components that could be hidden in the noise, we also examined 
all the spectra after smoothing them by a factor of 3. No broad and shallow 
components could be discerned. 

The only other available information relevant to the effective $b$ values 
for the DLAs in our sample is from the H I 21 cm absorption 
lines (Chengalur \& Kanekar 1999; Kanekar \& Chengalur 2001) and/or from 
lines of Ca II or Mg II seen in our MMT data for these systems 
(Khare et al. 2004). 
While the latter data cannot be used to estimate the $b$-values of individual 
components accurately, they provide  
upper limits to the effective $b$-values, and indicate a spread in 
velocities of several tens of km s$^{-1}$. Further details on the 
$b$ values are provided in discussions of the 
individual systems in section 3. The total column densities obtained 
with the present resolution are expected to be close to the actual values 
if very 
strongly saturated components are not present (Jenkins 1986). It seems unlikely 
that any of our Zn II, Cr II, or Fe II lines consist of single components 
with $b$-values less than 5 km s$^{-1}$. Indeed, $b$ values 
$\le 5 $ km s$^{-1}$ for the dominant component contributing to the Zn II 
lines are found in none of the eight DLAs at $z < 1.5$ and only 
$\approx 20 \%$ of the 24 DLAs at $z < 3.4$ for which profile fits from 
high-resolution data are available in the literature (Lu et al. 1995; Meyer 
et al. 1995; Prochaska \& Wolfe 1997; Lopez et al. 1999; Pettini et al. 1999, 2000; 
de la Varga et al. 2000; Molaro et al. 2001; Ellison \& Lopez 2001;  
Ledoux et al. 2002; Levshakov et al. 2002; Peroux et al. 2002; Ledoux et al. 
2003; Dessauges-Zavadsky et al. 2003, 2004). Thus, it would be 
unusual, although not impossible, in a sample of four low-redshift DLAs, 
to find two or more with $b < 5$. [Of course, it is possible that 
the low incidence of systems with $b < 5$ km s$^{-1}$ is a reflection of the 
limited resolution ($\approx$ 6-8 km s$^{-1}$ at best) achieved so far for DLA spectra 
with current instrumentation.  In principle, if DLAs often contain 
significant narrow, unresolved components, then many existing column 
density estimates for heavy elements in DLAs could be too low.] It would be  
desirable to have higher resolution observations in the future 
to understand the detailed component structure. This would need 
a large amount of observing time on a space-based telescope.  

In order to evaluate the effect of continuum placement uncertainties on our 
column density estimates, we slid the
continuum up and down by 0.3 times the 1 $\sigma$ noise level and measured the column
densities for these extrema by refitting the lines with FITS6P.  This was then 
used to calculate the uncertainty in the column densities 
due to the continuum placement uncertainty, which was then added in quadrature
with the errors associated with the photon noise, uncertainty in $b$-values,   
and uncertainty in component velocities to derive the final errors. 

Finally, we also estimated the column densities using the 
apparent optical depth (AOD) method (see, e.g., Savage \& Sembach 1991). 
The column densities obtained using this method were consistent 
with those obtained from profile fitting within the mutual 1 $\sigma$ uncertainties. 
However, the AOD results are expected to be less accurate 
than the profile-fitting approach for our data  
because of the relatively low SNR and resolution. This is 
because the AOD method weights all points across the observed 
profile equally, i.e., it does not take into account the fact that the observed 
profile ``should'' resemble the instrumental profile if the lines are 
unresolved or marginally resolved. Thus, for example, 
outlying downward noise points could spuriously increase the
computed AOD column density. We thus adopted the profile-fitting estimates 
of the Zn II, Cr II, and Fe II column densities and combined these with 
the known H I column densities to estimate the element abundances.  
For reference, we adopted the solar abundances recommended by Lodders (2003). 

\section{Results}

\subsection{Q0738+313} This quasar ($z_{em} = 0.635$) has two strong 
DLAs at $z = 0.0912$ and 
$z = 0.2212$ with log $N_{\rm H I} = 21.18^{+0.05}_{-0.06}$ and 
log $N_{\rm H I} = 20.90^{+0.07}_{-0.08}$, respectively (Rao \& Turnshek 
1998). The $z=0.2212$ 
system was known to be an Mg II absorber and identified as a DLA 
on the basis of {\it HST} UV spectra (Rao \& Turnshek 1998). The $z=0.0912$ 
absorber was discovered serendipitously in the same {\it HST} UV data. The 
quasar also shows 21-cm 
absorption at the redshifts of both the DLAs (Lane et al. 1998; 
Chengalur \& Kanekar 1999). 
The Zn II 
and Cr II lines of both systems lie outside the Ly-$\alpha$ forest. 
The wavelength coverage achieved with three separate grating settings 
included all five lines of Zn II and Cr II for each 
DLA.  The 
``unidentified''  
features near 2155, 2187, and 2217 {\AA} are likely to be C IV $\lambda$ 1548 
lines at $z = 0.392$, 0.413, and 0.432, respectively, with the corresponding 
$\lambda 1551$ features lost in the noise.

\subsubsection{The $z=0.0912$ DLA} 
Fig. 6 shows the velocity profiles of the Fe II, Zn II, and Cr II 
data for this DLA. The zero point of the velocity scale 
refers to the absorption redshift $z=0.0912$ inferred from the 21-cm and 
optical data. Counting both 
photon noise and continuum uncertainties, our data show 
weak $\approx 2.8 \, \sigma$, 2.7 $\sigma$, and 3.3 $\sigma$ features, 
respectively, at the wavelengths of the Zn II $\lambda 2026$, Cr II $\lambda 2056$, and Fe II 
$\lambda 2260$ lines. No other lines 
of these species were detected in our {\it HST} spectra. The 21-cm 
absorption in this object has been observed to be dominated by a single 
component at $z = 0.09118 \pm 0.00001$ with a 
full width at half maximum (FWHM) of $\sim 5$ km s$^{-1}$, 
indicating a Doppler parameter $b_{\rm {21 cm}}= 3.0$ km s$^{-1}$ 
for cold, neutral hydrogen 
(Chengalur \& Kanekar 1999). {\footnote{The Zn II, Cr II, 
and Fe II lines can arise in warmer H I gas or even H II regions, 
and could thus have effective $b$-values different from 
$b_{\rm {21 cm}}$. (If H II regions are present, the derived abundances 
[X/H] would, strictly speaking, be upper limits.)}}. Based on our
 complementary MMT spectra (Khare et al. 2004), we derive effective 
$b < 42$ km s$^{-1}$ for Ca II. 

We first carried out profile fits using only the potentially detected lines 
Zn II $\lambda$ 2026, Cr II $\lambda$ 2056, and Fe II $\lambda 2260$.  The fits  
were then repeated by fitting these lines together with the undetected lines 
Zn II $\lambda$ 2062; Cr II $\lambda \lambda 2062, 2066$; and Fe II $\lambda 2249$. 
To be conservative, we adopt the results from these latter fits, and list the derived 
Zn II, Cr II, and Fe II column densities, and the corresponding abundances in Table 5. 
The error bars on column densities and abundances for these fits 
include uncertainties 
from Poisson statistics and uncertainties in $b$ value, 
central velocity, and continuum placement. 
The formal estimate of the best fit Zn II column density in this DLA has a relatively large 
total error, therefore we list it only as a 3 $\sigma$ upper limit. 
To explore the dependence 
of the estimated column densities on the $b$ values, we also performed 
single-component fits to the individual detected lines for several (fixed) 
choices of $b$, i.e., 5, 10, 15, and 20 km s$^{-1}$, which are also 
listed in Table 5. The column densities for these single line fits with $b$ 
fixed are higher than that for the multiple-line fit with $N, b, v$ varied, 
because the former included fits to only detected lines, while the 
latter included fits to both detected and undetected lines. 

The dashed curves in Fig. 6 show the profiles for the best-fit $N$, $b$, 
and $v$ parameters obtained by varying these parameters simultaneosuly 
for the multiple lines stated above. The estimate of $b$ value obtained in this way 
is $11.2 \pm 6.7$ km s$^{-1}$, where the uncertainties in $b$ were 
calculated simultaneously with the uncertainties in $N$ and $v$.

For pictorial illustration, we created simulations of the Zn II 
$\lambda$ 2026  line for solar, 32 $\%$ solar, and 10 $\%$ solar metallicities 
([Zn/H] = 0.0, -0.5, and -1.0), assuming the best-fit $b$ and $v$ values,  
$b=11.2$ km s$^{-1}$ and $v=9.9$ km s$^{-1}$. 
These simulations are shown in Fig. 7, overlaid on top of our data, and  
support the profile fitting results that the metallicity appears to be low. 

\subsubsection{The $z=0.2212$ DLA} This DLA has also been detected in 21-cm 
absorption. Kanekar, Ghosh, \& Chengalur (2001) 
have reported three components in the 21-cm absorption line, 
the dominant one having FWHM = 18.6 
km s$^{-1}$ and comprising $\sim 87 \%$ of the H I column density. 
This indicates a Doppler parameter $b_{\rm {21 cm}}= 11.2$ km s$^{-1}$ 
for most of the cold, neutral hydrogen. 
Our MMT data suggest effective $b \le 38$ km s$^{-1}$ for Mg I and 
Mg II in this DLA (Khare et al. 
2004).

Fig. 8 shows the velocity profiles of the Zn II and Cr II data for this DLA. 
The zero point of the velocity scale refers to the absorption redshift 
$z=0.2212$ inferred from the 21-cm and optical data. 
Our data show $\approx$ 2.8 $\sigma$ and 3.5 $\sigma$ 
features at the wavelengths of the Cr II $\lambda$ 2062 and Zn II $\lambda$ 
2062 lines. These lines lie in the C IV  broad 
emission line of the quasar, and therefore have  higher continuum SNR 
than that for the Zn II $\lambda 2026$ and Cr II $\lambda 2056$ lines 
which are not detected. 

We first carried out profile fits using the potentially detected lines 
Zn II $\lambda$ 2062 and Cr II $\lambda$ 2062 together with the undetected line  
Cr II $\lambda 2066$. The undetected Zn II $\lambda$ 2026 and Cr II $\lambda 2056$ lines 
were not included in the fits because they lie in regions of the spectrum with lower 
continuum signal level. The estimate of the $b$ value obtained in this way 
is $12.2 \pm 7.7$ km s$^{-1}$. In Table 6, we list 
the derived Zn II and Cr II column densities, and the corresponding 
abundances. 
The dashed curves in Fig. 8 show the profiles for these best-fit 
$N$, $b$, and $v$ parameters. Also listed in Table 6 are results of 
single-component fits to the individual detected lines for several (fixed) 
choices of $b$, i.e., 5, 10, 15, and 20 km s$^{-1}$. The column densities 
of Cr II for these single line fits with $b$ 
fixed are higher than that for the multiple-line fit with $N, b, v$ varied, 
because the former included only the detected lines (Cr II $\lambda 2062$, 
Zn II $\lambda 2062$), while the latter also included the undetected Cr II 
$\lambda 2066$ line. 

We next checked whether the Zn II and Cr II column densities thus estimated are 
consistent with the lack of significant features at the positions of the 
Zn II $\lambda 
2026$ and Cr II $\lambda 2056$ lines in the noisier parts of our spectra. To 
do this, we generated simulated Zn II $\lambda 2026$ and Cr II $\lambda 2056$ 
lines with the best-fit parameters estimated from the 
Cr II $\lambda \, \lambda$ 2062, 2066 and Zn II $\lambda$ 2062 lines. At our 
resolution, the Zn II $\lambda 
2026$ and Cr II $\lambda 2056$ features would be 7 pixels wide, but are 
lost in the noise. 
The 3 $\sigma$ limits on the rest-frame equivalent width over these 7-pixel 
wide simulated features are about 89 m{\AA} and 75 m{\AA}, 
respectively, at the positions of the  Zn II $\lambda$ 2026 and 
Cr II $\lambda$ 2056 
lines. For the best-fit $b$ value of 12.2 km s$^{-1}$, 
this implies 3 $\sigma$ upper limits of 
log $N_{\rm {Zn II}} < 12.81$, 
i.e. [Zn/H] $< -0.72$, and log $N_{\rm {Cr II}} < 13.37$, 
i.e. [Cr/H] $< -1.18$. These limits are consistent with 
the values obtained from the data for Zn II $\lambda 2062$ 
and Cr II $\lambda \, \lambda 2062, 2066$. Thus  
the Zn abundance in this system is likely to be close to -0.70, 
the value obained from the $\lambda 2062$ line. 
Nevertheless, to be conservative, we adopt 
an upper limit [Zn/H] $\le -0.70$ for this system.

For pictorial illustration, we show in Fig. 9 simulations of the Zn II 
$\lambda$ 2062 line for solar, 32$\%$ solar, and 10$\%$ solar metallicities 
([Zn/H] = 0.0, -0.5, and -1.0), assuming the best-fit $b$ and $v$ values,  
$b=12.2$ km s$^{-1}$ and $v=15.3$ km s$^{-1}$.  
 The parameters of the Cr II 
$\lambda$ 2062 line are held fixed in these simulations, as determined from 
the best-fit obtained by fitting together the 
Cr II $\lambda \, \lambda$ 2062, 2066, Zn II 
$\lambda$ 2062 lines. These simulations   
support the profile fitting results that the metallicity appears to be low. 

\subsection{Q0827+243} 
This quasar ($z_{em} = 0.939$) has a DLA at $z = 0.5247$ with 
log $N_{\rm H I} = 20.30^{+0.04}_{-0.05}$ (Rao \& Turnshek 
2000). The system was known 
from ground-based studies as a strong Mg II absorber at $z = 0.5247$, 
and found to be a DLA in {\it HST} UV spectra (Rao \& Turnshek 2000). 
The quasar also shows a relatively broad and possibly multi-component 
21-cm absorption feature at 
$z = 0.52476 \pm 0.00005$, with FWHM $\sim 50$ km s$^{-1}$ , indicating 
$b_{\rm 21 cm} \sim 30$ km s$^{-1}$ in the cold neutral hydrogen (Kanekar \& 
Chengalur 2001). Our MMT data show Fe II and Mn II absorption 
at $z = 0.5249$, with an effective $b \le 35$ km s$^{-1}$ 
(Khare et al. 2004). 

Unfortunately, this quasar was in a significantly fainter state at the time of 
our {\it HST} observations than expected from previous UV data. Therefore the 
SNR in our {\it HST} data was poorer than expected. The wavelength region covered 
all five lines 
of Zn II and Cr II, but no lines could be detected. Fig. 10 shows the 
velocity profiles for this DLA. The zero point of the velocity scale 
refers to the absorption redshift $z=0.5247$ inferred  
from the 21-cm and optical data. Although weak features are seen at  
-20 km s$^{-1}$ in the Cr II $\lambda$ 2062 line and possibly 
at -40 km s$^{-1}$ in the Zn II $\lambda 
2062$ line, they are comparable to the noise. Furthermore, 
there is no evidence for these features in the stronger Zn II $\lambda$ 
2026 or Cr II $\lambda$ 2056 lines. The 3 $\sigma$ rest 
frame equivalent width limits in our {\it HST} data at the expected positions of the 
Zn II $\lambda$ 2026 and Cr II $\lambda$ 
2056 lines are 96 m{\AA} and 90 m{\AA}, respectively. The corresponding 
3 $\sigma$ upper limits on the Zn II and Cr II column densities and abundances 
are listed in Table 7 for $b = 5, 10, 15,$ and 20 km s$^{-1}$. The differences 
between the limits for $b = 15$ and 20 are small, so the limits 
for higher $b$-values are expected to be close to those for $b = 20$ 
km s$^{-1}$. For example, for $b = 30$, the $b$ value of the 21-cm absorption 
line, the limits are about 0.1 dex lower 
than for $b = 10$. The dashed curves in Fig. 10 show the profiles corresponding 
to the 3 $\sigma$ upper limits for $b=10$.

The low SNR of our STIS data for Q0827+243 prevents us from putting tight constraints on 
the Zn and Cr abundances in the DLA absorber along this sightline.  
We note that based on MMT spectra of this target, the abundance  
of Fe in this DLA is  [Fe/H] $= -1.02 \pm 0.05$ (Khare et al. 2004). The 
total (gas + solid phase) metallicity could of course be higher. 
But the DLA toward this quasar has a 
much lower $N_{\rm H I}$ than that of the other three DLAs in our sample. 
Therefore, unless 
the metallicity of this DLA is substantially supersolar (which would imply 
a very large amount of dust depletion), it would not significantly affect 
our estimates of the global metallicity of DLAs at low redshifts, which 
anyway treat this DLA as an upper limit.  

\subsection{Q0952+179} This quasar ($z_{em} = 1.472$) has a strong DLA 
at $z=0.2378$ with log $N_{\rm H I} = 21.32^{+0.05}_{-0.06}$ (Rao \& Turnshek 
2000). The absorber 
was known to be a strong Mg II system and found to be a DLA in {\it HST} UV 
spectra (Rao \& Turnshek 2000). The quasar also shows 21-cm absorption 
at the same redshift (Kanekar \& Chengalur 2001). 

Our wavelength setting covered the Zn II $\lambda$ 2062 and Cr II 
$\lambda \, \lambda$ 2056, 2062, 2066 lines. A separate setting to cover only 
the 
Zn II $\lambda$ 2026 line was not obtained in view of the large number of 
orbits needed for this faint quasar to cover even a single grating setting to a 
reasonable SNR. Tables 3 and 4 list the lines measured in the final spectrum. 
Fig. 11 shows the velocity profiles for the Zn II and Cr II lines 
in this absorber. The zero point of the velocity scale 
refers to the absorption redshift $z=0.2378$ inferred from the 21-cm and optical data. 
Our data show weak 2.5 $\sigma$, 4.2 $\sigma$, 
and 2.1 $\sigma$ 
features at the wavelengths of the Cr II $\lambda$ 2056, Cr II $\lambda$ 2062, 
and Zn II $\lambda$ 2062 lines. The strong feature appearing near 200 km 
s$^{-1}$ in the Cr II $\lambda$ 2056 line cannot be Cr II, because 
it is not seen in the Cr II $\lambda 2062$ line (which is lower in 
$f$ value by a factor of only 1.3) even though the SNR in the continuum 
is comparable. Furthermore, the velocity of this component would be too 
high compared to the redshifts of the Mg II and 21-cm 
absorption lines. We take this feature to be a Lyman-$\alpha$ forest line. 

The only other metal lines available in the literature for this DLA are the 
Mg II $\lambda \, \lambda$ 2796, 2803 lines at $z = 0.2378$ in lower 
resolution spectra. The 21-cm absorption line 
detected by Kanekar \& Chengalur (2001) at $z = 0.23780 \pm 0.00002$ shows a 
single component with a FWHM of 7.7 km s$^{-1}$, indicating 
$b_{\rm {21 cm}}$ = 4.6 km s$^{-1}$ for cold neutral hydrogen. 

We carried out profile fits using simultaneously all the lines of Zn II and Cr II that 
were included in our data, i.e.,  Zn II $\lambda$ 2062 and Cr II $\lambda \lambda$ 
2056, 2062, 2066. The dashed curves in Fig. 11 show the profiles for the best-fit 
$N$, $b$, and $v$ parameters. The estimate of the $b$ value obtained from the fits  
is $15.0 \pm 7.7$ km s$^{-1}$. In Table 8, we list 
the derived Zn II and Cr II column densities, and the corresponding 
abundances. 
The formal estimate of the best fit Zn II column density in this DLA has a 
relatively large 
total error; therefore we list it only as a 3 $\sigma$ upper limit. 
Also listed in Table 8 are results of 
single-component fits to the individual detected lines for several (fixed) 
choices of $b$, i.e., 5, 10, 15, and 20 km s$^{-1}$. 

For illustrative purposes we show, in Fig. 12, simulations of the Zn II 
$\lambda$ 2062 line for solar, 32$\%$ solar, and 10$\%$ solar metallicities 
([Zn/H] = 0.0, -0.5, and -1.0), assuming the best-fit $b$ and $v$ values,  
$b=15.0$ km s$^{-1}$ and $v=8.7$ km s$^{-1}$. 
The parameters of the Cr II 
$\lambda$ 2062 line are held fixed in these simulations. These simulations   
support the profile fitting results that the metallicity appears to be low.

Our results are summarized in Table 9. The errorbars on the abundance estimates 
include photon noise and the uncertainties in  $b$ value,  
central velocity, and continuum placement. The [Zn/H] 
value for the DLA in Q0827+243 is not well constrained 
because of the very low SNR in the spectrum of this quasar. The [Zn/H] 
values appear to be low for the DLAs toward Q0738+313 and Q0952+179. 

\section{The Metallicity-Redshift Relation for DLAs}

To examine the implications of our data for the mean metallicity-redshift relation 
of DLAs, we combined our data with other Zn measurements at higher redshifts. 
While doing this, we included only the ``classical DLAs'', i.e., 
systems with log $N_{\rm H I} \ga 20.3$. 
Furthermore, we excluded systems with radial velocities within 
3000 km s$^{-1}$ from the quasar 
emission redshifts, since these associated systems may not have similar 
abundances to those of the intervening systems at the same redshifts. 
The combined sample contains a total of 87  
DLAs with Zn detections or limits, in the redshift range $0.09 < z < 3.90$, 
and is summarized in Table 10.  The sample includes our {\it HST} data, 
MMT data for 4 intermediate redshift DLAs recently obtained by us  
(Khare et al. 2004), and other 
data from the literature, as detailed in Table 10. While combining 
 these measurements from different studies, we have normalized them all 
 to the same set of oscillator strengths and solar abundances that we 
 have used for our {\it HST} data. Our {\it HST} observations  
have provided 3 new Zn measurements at $z < 0.39$ where no data existed 
before. 
This sample, together with our recent data at $0.6 < z < 1.5$ (Khare et al. 2004), 
has doubled the $z < 1.5$ sample and increased the $z < 1$ sample by a factor 
of 2.5. 
Most of the other recent observations have increased the samples 
at $z \gtrsim 2$. 

We now examine the evolution of the global Zn metallicity for these 87 DLAs.  
Of these 87 systems, 51 are detections and 36 are limits. One way to deal 
with the limits is to consider that the true Zn column density for a system 
with a limit is somewhere between zero and the upper limit value. 
With this in mind, we first constructed two extreme samples, a ``maximum-limits'' 
sample, where we treated the Zn limits as detections, and a ``minimum-limits'' 
sample where we treated the Zn limits as zeros. 
For any individual system, these cases cover the full range of possible 
values the Zn column densities can take. 
For the sake of pictorial illustration, we divided each 
of these samples into six redshift bins with 14 or 15 absorbers each. Figure 13 shows 
the logarithm of the $N({\rm H \, I})$-weighted mean metallicity 
relative to the solar value 
for these six bins, plotted as a function of redshift. Each point 
is plotted at the median redshift in its bin. Horizontal 
bars for each bin extend from the lowest DLA redshift to the highest 
DLA redshift within that bin. 
The top panel shows the ``maximum-limits'' case. The filled circles  
in the bottom panel show the 
``minimum-limits'' case with limits treated as zeros. The vertical error bars 
in both panels show the 1 $\sigma$ uncertainties in the $N_{\rm H I}$-weighted mean 
metallicities, and include both the measurement uncertainty in the individual 
Zn II and H I column density measurements, and the sampling uncertainty. 
Further descriptions of the statistical procedures can be found 
in Kulkarni \& Fall (2002). Clearly, Fig. 13 shows that the 
$N_{\rm H I}$-weighted mean metallicity of DLAs does not rise fast enough 
to reach solar or near-solar metallicities at $z = 0$. 

The above approach is perfectly adequate for low redshifts, which is the 
main redshift range of interest for our paper, since the number of Zn limits 
is relatively small at low redshifts. However, at high 
redshifts, there are a large number of Zn limits. At the same time, 
for most high-$z$ systems with Zn limits, measurements of other elements are 
available. We therefore examined how much the results of the ``minimum-limits'' 
sample would change if we use information available from other elements for 
the systems with Zn limits. 
To do this, we created a third sample, using  
conservative constraints on the metallicities implied by the information 
for the other elements (detections or lower limits) available in the 
literature (Lu et al. 1996; Ledoux et al. 1998; Ledoux, 
Bergeron, \& Petitjean 2002; Prochaska \& Wolfe 1999; Prochaska 
et al. 2001a; 2002; 2003a; 2003b; 2003c; and references therein). For 23 of 
the 36 systems with Zn upper limits, we estimated lower limits 
for Zn using measurements of the weakly depleted element Si, and 
assuming the total (gas + solid phase) relative abundance Si/Zn 
to be in solar proportion. For one system, we set the Zn abundance 
to the abundance of the undepleted element S, assuming solar S/Zn ratio. 
For most of the remaining systems without Si, S, or O measurements, we 
set the constraints on Zn using measurements of Fe (8 systems) 
or Cr (2 systems). While setting the Zn constraints using Fe or Cr, we 
conservatively added 0.4 dex 
to correct for a typical amount of depletion in DLAs, unless other elements 
suggested no depletion. Finally, for two systems, no other measurements 
could be used to set a lower limit on Zn. In these cases we set the lower 
limits on Zn abundances to 0.01 solar values, since DLA metallicities 
are seldom found to be below 0.01 solar. The unfilled circles  
in the bottom panel of Fig. 13 show the mean metallicity in each 
redshift bin for this 
modified minimum-limits sample with the information 
on other elements used in cases of Zn limits. The two bins at $z > 2.8$  
show some change as a result of 
using the other elements, but there is very little or no change in the 
other four bins at $z \la 2.8$. Thus our conclusion about the low mean metallicity 
at low redshifts is not sensitive to whether or not information from 
other elements is used in cases of Zn limits. 

We now quantify the shape of the metallicity-redshift relation using 
several statistical tests outlined in Kulkarni \& Fall (2002). 
Table 11 lists the results for the binned 
metallicity-redshift relation for the three samples considered above, 
in columns 4, 5, and 6. 
The logarithm of the $N_{\rm H I}$-weighted mean 
metallicity in the lowest redshift bin $0.09 < z < 1.37$ is $-0.86 \pm 0.11$ 
for the maximum-limits sample, $-1.01 \pm 0.14$ for the minimum-limits 
sample using Zn only (i.e. treating Zn limits as zeros), and $-0.93 \pm 0.10$ 
for the minimum-limits 
sample using information from other elements in cases of Zn limits. 
For the $ z < 0.6$ range not accessible to ground-based 
telescopes, the logarithm of the $N_{\rm H I}$-weighted mean 
metallicity is -0.78 for the maximum-limits sample and 
-0.99 for the minimum-limits  sample using information from other elements 
in cases of Zn limits. 

We next fit the data with a straight line relation of the form
${\rm log} \,({\overline Z / Z_{\odot}}) = 
{\rm log} \,({\overline Z_{0}/ Z_{\odot}}) + Bz$, and list the 
results for the parameters in Table 12. 
For the binned data (with six bins in the range $0.09 < z < 3.90$), the 
linear regression slope $B$ of the logarithm of the 
metallicity vs. redshift relation is $-0.18 \pm 0.06$ for the maximum-limits  
sample, $-0.22 \pm 0.08$ for the minimum-limits sample with Zn only 
(i.e. with Zn limits treated as zeros), and $-0.23 \pm 0.06$ for the minimum-limits sample 
using constraints from other elments in cases of Zn limits. 
The linear regression intercept of the 
metallicity-redshift relation (which corresponds to predicted present-day 
metallicity) is $-0.74 \pm 0.15$ for the maximum-limits  
sample, $-0.75 \pm 0.18$ for the minimum-limits sample with Zn only,
 and $-0.71 \pm 0.13$ for the minimum-limits sample using constraints 
from other elments in cases of Zn limits. 
Of course, as mentioned by Kulkarni \& Fall (2002; see their 
footnote 3), unlike the means in individual bins, the slopes and 
intercepts thus determined do not necessarily bracket the true ranges 
of the slope and the intercept. This is because, in principle, 
all the true values in cases of upper limits could be closer to the limits 
at high redshifts, and closer to zero at low redshifts. 
Nevertheless, these cases are indicative 
of the ranges the slope and the intercept are likely to take. 

Another approximate, although not rigorously justified, way 
to get an indication of the slope and intercept of the metallicity-redshift 
relation from our maximum and minimum limits cases is to take the mean of the 
logarithmic metallicity estimates for the maximum limits 
and minimum limits cases in 
each redshift bin. Column (7) of Table 11 lists the means of the 
logarithmic metallicity estimates for the 
maximum limits case (column 4) and the minimum limits case with information from 
other elements in cases of Zn limits (column 6). 
The error bars on these mean values include 
the uncertainty in the values for the maximum and minimum limits cases 
and the standard deviation of these two values with respect to 
their mean value. The error bar on the mean value is large for the 
highest redshift bin ($3.28 < z < 3.90$), which contains mostly limits. 
The linear regression 
slope and intercept of the global metallicity-redshift relation 
obtained on using these mean logarithmic metallicity 
estimates in each bin 
are $-0.20 \pm 0.07$ and $-0.72 \pm 0.16$, respectively. 

We have also repeated several other calculations of the metallicity-redshift 
relation excluding the highest redshift bin (where there are 
mostly limits rather than detections of Zn); using an unbinned 
$N({\rm H \, I})$-weighted nonlinear $\chi^{2}$ approach; or 
using survival analysis, following the 
methodology of Kulkarni \& Fall (2002). 
In the survival analysis calculations, we computed the Kaplan-Meier 
(K-M) estimator, using the astronomical survival analysis package 
ASURV rev. 1.2 (Isobe \& Feigelson 1990) which implements the methods 
of Feigelson \& Nelson (1985). The error bars on the survival analysis 
estimates include the sampling and measurement uncertainties in the 
column densities. 
The results of all these calculations are listed in Tables 11 and 12. 

Finally, to assess the implications of our {\it HST} data alone for the 
statistics of the metallicity-redshift relation, 
we also repeated our analysis for the 83 systems left after taking out 
the other new data used here for 4 intermediate-redshift systems from 
Khare et al. (2004). These calculations yielded essentially identical 
estimates of the metallicity-redshift 
relation. For example, for the maximum-limits sample, the 
$N_{\rm H I}$-weighted mean metallcity in the lowest redshift bin  
changes by only $\approx 0.03$ dex. The intercept and slope of 
the metallicity-redshift relation also change by only $\la 0.03$. 

The various tests described above give slightly different values, but they all 
confirm that the slope of the metallicity-redshift relation is small, 
but probably not zero, and the intercept  
is low. In other words, the global metallicity of 
DLAs does not seem to rise to the solar value at low redshifts. 
This just confirms the visual impression from Fig. 13. 

Our estimates of the metallicity-redshift relation are 
consistent with previous results of Kulkarni \& Fall (2002) and Prochaska 
et al. (2003b). We have not 
included metallicity estimates based 
on X-ray absorption in our analysis here because the X-ray data are 
expected to reflect mainly 
the abundance of O, and there could be a systematic difference between 
[Zn/H] and the [X/H] derived from the X-ray absorption data. 
{\footnote{Prochaska et al. (2003b) reported a slope of $-0.25 \pm 0.07$ 
for the metallicity-redshift relation in the redshift range $0.5 < z < 4.7$., a detection 
of evolution in the global mean metallicity at $ \ga 3 \sigma$ level.  
Their analysis included the X-ray measurement  
from Junkkarinen et al. (2004) for the $z = 0.52$ DLA toward Q0235+164. 
This system, because of its low redshift and very high H I column density 
(log $N_{\rm H I} = 21.80$), 
has a very strong influence on the mean metallicity in the lowest redshift 
bin, and therefore on the overall slope of the mean metallicity-redshift 
relation. 
When Kulkarni \& Fall (2002) included X-ray absorbers in their 
analysis, they also found evolution at $\ga 3 \sigma$ level (see 
their foonote 4). For most of their analysis, Kulkarni \& Fall (2002) 
left out X-ray absorbers since the X-ray measurements primarily 
trace the O abundance, which could differ systematically from the Zn 
abundance. }} 

To summarize, our analysis suggests that the global metallicity of DLAs 
stays low at low redshifts, and evolves slowly at most.  
We stress, however, that the current samples are still small, and 
further observations, 
particularly at the low and intermediate redshifts are necessary to 
decrease the uncertainties in the shape of the metallicity-redshift relation. 

We now compare our results with some models of cosmic chemical evolution. 
The short-dashed, long-dashed, solid, and short-dash-long-dashed curves in Figure 13 
show, respectively, (a) the mean interstellar metallicity in the closed-box 
and outflow models of Pei \& Fall (1995); 
(b) the mean Zn metallicity in the Malaney \& Chaboyer (1996) model with 
a constant Zn yield, a slope of $-1.70$ for the initial mass function 
(IMF), and the evolution of the neutral gas density 
as computed by Pei \& Fall (1995); (c) the mean interstellar 
metallicity in the Pei et al. (1999) model with the optimum fit for the 
cosmic infrared background intensity; and 
(d) the mean metallicity of cold interstellar gas in the 
semi-analytic model of Somerville et al. (2001). 
Models (a) and (c) agree with the data at high redshifts, but evolve 
too rapidly toward solar metallicity to be consistent with the data 
at low redshifts. Models (b) and (d) have 
lower slopes but predict too high a metallicity at all redshifts. 
None of the models agrees completely with all the observations.

\section{Discussion}
The main new result from our study is that, in all the 3 DLAs for which our 
observations offer meaningful constraints on [Zn/H], the metallicities appear 
to be substantially sub-solar, and that there is at most only a 
weak evolution in the global mean metallicity with redshift. To understand 
this puzzling result, we now explore (a) whether our results are affected 
in any way by the use of Zn as a metallicity indicator, and 
(b) how our results fit in with the morpohological information available 
for these DLAs. Finally, we discuss the role of 
potential selection effects caused by dust obscuration in the DLAs.

\subsection{Zinc as the Metallicity Indicator}

Could the apparent low metallicity for most of our target DLAs be an 
artifact of the use of Zn as the metallicity indicator? To explore this 
question, we now examine the advantages and disadvantages of using Zn. 

The conclusion that [Zn/Fe] $\sim 0$ in Galactic halo and disk stars 
for metallicities [Fe/H] $\ga -3$ 
that followed from earlier work (Sneden, Gratton, \& Crocker 1991) 
has been strongly supported by recent studies (e.g., Mishenina et al. 2002; 
Cayrel et al. 2004; Nissen et al. 2004). There is some evidence 
for a small overabundance of Zn by $\sim 0.1-0.2$ dex in metal-poor disk stars 
and halo stars with [Fe/H] $< -2$ (e.g., Prochaska et al. 2000; 
Bihain et al. 2004; Cayrel et al. 2004). Chen et al. (2004) have also 
recently reported [Zn/Fe] $\sim 0.1-0.15$ dex in disk stars with [Fe/H] 
$> -0.9$. However, in all of these studies, there are systematic uncertainties  
in [Zn/Fe] at the level of $\approx \pm 0.1$ dex, and overall Zn 
tracks Fe quite closely for [Fe/H] $> -2$. These observations 
are not completely 
understood in nucleosynthetic models of massive stars. The models are 
complex and the value of [Zn/Fe] depends on several parameters such 
as the position of the mass cut, 
the explosion energy, the degree of neutronization, and the consequences of a 
neutrino-driven wind (Woosley \& Weaver 1995; Chieffi \& Limongi 2002; 
Umeda \& Nomoto 2002; Woosley, Heger, \& Weaver 2002; Fenner, Prochaska, \& 
Gibson 2004). Despite 
these uncertainties in the theoretical understanding, the Galactic 
stellar data support the use of Zn as a tracer of metallicity. 

There are some observational difficulties in measuring the 
Zn abundances in DLAs. The Zn II $\lambda \lambda \, 2026$, 2062 lines are 
weak and sometimes hard to detect. On the other hand, for 
that same reason, they are usually unsaturated. At $z \ga 3$, the Zn II 
lines lie in 
spectral regions affected by night sky lines, and thus become 
difficult to detect. At the same time, the long wavelengths of the Zn II lines 
also make them accessible in near-UV and optical spectra for the range 
$z \la 3$ which spans $\approx 90 \%$ of the cosmic history. 

While some studies have suggested that ionization corrections for Zn 
could be significant (e.g., Howk \& Sembach 1999), others have given 
convincing arguments that such corrections are negligible. For example, 
Vladilo et al. (2001) have pointed out that the ionization corrections 
are uncertain because of uncertainties in the Zn recombination coefficients 
and ionization cross sections. They have provided evidence 
that the Zn ionization corrections could be lower, and concluded that DLA 
metallicity estimates based on [Zn/H] are unlikely to be significantly 
affected by ionization effects. Indeed, at the large H I column densities 
(log $N_{\rm H I} = 20.9-21.3$) of the 3 DLAs in our sample where we find low [Zn/H], the 
ionization correction for [Zn/H] is expected to be at most about 0.3 dex, 
and probably smaller for more accurate atomic parameters  
(see, e. g., Fig. 16 of Vladilo et al. 2001). 

Thus, while Zn, like all other elements, has both 
advantages and disadvantages, it is still an appropriate choice as a metallicity 
indicator in DLAs (see, e.g., Bihain et al. 2004; Nissen et al. 2004). 
Furthermore, the various uncertainties mentioned above are not large 
enough to make significant difference to our results regarding a low global 
mean metallicity ($\approx$ -0.7 to -1.0 dex) at low redshifts. Overall,
 we conclude 
that our main result is not affected much by the use of Zn as the 
primary metallicity indicator. 
 
\subsection{Comparison with Morphological Information}

Could the low metallicities in our low-redshift DLAs be understood 
in terms of the nature of the underlying galaxies? 
In fact, the low metallicities are consistent with the 
morphological information available on the candidate galaxies 
believed to give rise to these DLAs. 

The $z=0.0912$ DLA toward Q0738+313 has been tentatively identified as associated 
with fuzz surrounding the quasar, and classified as a low surface brightness 
galaxy. This galaxy has an estimated luminosity of $0.08 L_{K}^{*}$ and impact parameter 
$ < 3.6 \, h^{-1}$ kpc for 
$\Omega_{m} = 1$, $\Omega_{\Lambda} = 0$, and $H_{0} = 65 \, h $ km s$^{-1}$ 
Mpc$^{-1}$ (Cohen 2001; Turnshek et al. 2001). The $z=0.2212$ DLA toward 
the same quasar has been spectroscopically identified as a dwarf spiral galaxy, 
with B and K-band luminosities of $0.1 L_{B}^{*}$ and $0.1 L_{K}^{*}$, 
respectively, and impact parameter of 20 $h^{-1}$ kpc (Cohen 2001; 
Turnshek et al. 2001). 

From ground-based images in the 
B, R, J, and K bands, Rao et al. (2003) have suggested that 
the best candidates for the $z=0.2378$ DLA toward 
Q0952+179 are two dwarf low surface brightness galaxies with a combined luminosity 
of $0.02 L_{K}^{*}$, and impact parameter $ < 4.5 \, h^{-1}$ kpc. However, these candidate objects have not yet been confirmed 
spectroscopically. In any case, the DLA in this field cannot be an 
$L^{*}$ galaxy, as none of the luminous galaxies in the field have the same 
redshift as the DLA (Bergeron \& Boiss\'e 1991). 

For Q0827+243 ($z_{abs} = 0.5247$), where we have only a loose upper 
limit on the metallicity, a galaxy at $z=0.5258$ has been spectroscopically 
confirmed at an angular separation of 6", corresponding to a 
projected separation of 34 $h^{-1}$ kpc. 
This galaxy has B and K-band luminosities of $0.8 L_{B}^{*}$ and $1.2 L_{K}^{*}$, 
respectively. Rao et al. (2003) have suggested a ``disturbed spiral'' morphology 
for this galaxy.  Unfortunately, our Zn data do not yield a firm determination 
of the metallicity in this DLA. The 
Fe abundance is low ([Fe/H] $= -1.02 \pm 0.05$; Khare et al. 2004); but 
this could be partly because of depletion of Fe on dust grains, and the 
total (gas + solid phase) metallicity could be higher. In any case, 
at the large impact parameter for this absorber, the metallicity 
is likely to be low, for a typical abundance gradient of $\sim 0.05-0.1$ 
dex kpc$^{-1}$ observed in the Milky Way (e.g., Friel \& Janes 1993). 
 
\subsection{Potential Selection Effects}

Our observations of low metallicities in low-redshift DLAs are surprising 
because the mean mass-weighted interstellar metallicity of nearby galaxies 
is found to be 
close to solar (e.g., Kulkarni \& Fall 2002). Thus, samples derived from 
absorption-line studies of optically selected quasars seem to be biased toward   
less metal-rich galaxies than would be expected from 
an H I emission-selected sample of galaxies. 

The low metallicities of our low-$z$ DLAs also appear to contradict the 
expectations of a solar or near-solar metallicity at $z=0$ from most cosmic chemical 
evolution models based on the global star formation history of galaxies.  
Our observations thus suggest that the current DLA samples, especially those 
at low redshifts, might not be representative of the global galaxy population. 
If the observed DLAs have low metallicities at all redshifts, it is 
possible that the DLA galaxies 
are dominated by dwarf or low surface brightness galaxies. Some earlier 
studies have suggested that 
DLAs could arise in dwarf galaxies (e.g., York et al. 1986, 
Matteucci et al. 1997). Indeed, some 
imaging studies have suggested DLAs at low redshifts to be 
predominantly dwarf or low surface brightness galaxies 
(e.g., Rao et al. 2003). 
In fact, as discussed in section 5.2, three out of the four DLAs in our 
sample are tentatively identified as 
dwarf or low surface brightness galaxies (although only one of them is 
spectroscopically confirmed). Again, taken at face value, 
this is very surprising 
because the H I emission studies of galaxies at $z = 0$ show that 
most of the H I at the present epoch resides in large spirals 
(e.g., Rao \& Briggs 1993; Zwaan, Briggs, \& Sprayberry 2001). Of course, the 
identification of low-$z$ DLAs with dwarf or low surface brightness 
galaxies is not always certain, since many imaging 
studies (e.g., LeBrun et al. 1997; Rao et al. 2003) have not always 
confirmed the redshifts of the candidate galaxies. Chen \& Lanzetta 
(2003) have obtained more extensive photometric redshift data for 
DLA fields and suggested that the contribution of dwarf galaxies to 
 DLAs at $z < 1$ 
is not significant. Rosenberg \& Schneider (2003) have also suggested 
that the luminosity function of low-redshift DLAs may be fairly flat over 
several orders of magnitude. 
Thus, there seems to be no full consensus about the nature of low-redshift 
DLAs. Clearly, it is necessary to obtain further 
imaging studies and redshift identifications of low-redshift DLAs 
and take into account the luminosity uncertainties of the DLA galaxies. 

Our observations of low metallicities in low-redshift DLAs 
suggest that these galaxies have had less nucleosynthetic processing 
and hence low time-averaged star formation rates (SFRs). On the other hand, 
observations of the UV luminosity density of galaxies (e.g. Lilly et al. 
1996; Madau et al. 1996) indicate that the global average SFR in galaxies 
was high at $z \ga 1.5$. In fact, it is this high global SFR at high 
redshifts that 
drives the metallicity to rise up to solar or near-solar levels by the 
present epoch in most cosmic chemical evolution models.   
There are also indications from C II$^{*}$ absorption that some 
DLAs may have high SFRs and could be efficient producers of heavy 
elements (e.g., Wolfe et al. 2003). 
If the observed low-redshift DLAs are largely metal-poor, then where 
are the more metal-rich low-$z$ absorbers expected to have near-solar 
metallicities? 
Why are the absorption studies not detecting more luminous and 
metal-rich DLAs even at low redshifts? Such a selection effect could be 
caused by dust obscuration in the DLAs. This is because the DLA galaxies 
with the highest column densities of 
metals are likely to be those with the highest column densities of dust,
and these may obscure background quasars to such a degree that
some of them are omitted from optically selected samples (Fall 
\& Pei 1993; Boisse et al. 1998). As a result of obscuration, 
the observed mean metallicity in 
the DLA galaxies, and hence the data points in Figure 13, could  
lie systematically below the true mean metallicity. The model predictions 
plotted in Fig. 13, however, refer to the ``true'' mean metallicity (not 
corrected for dust obscuration). Indeed, if the true mean metallicity and 
dust content increase with decreasing redshift, more DLAs could be 
missing at lower redshifts. This would flatten the metallicity-redshift 
relation, as possibly shown by our observations. 

In principle, a comparison of metal abundances in DLAs from optical vs. 
radio-selected quasars can help to quantify the dust selection effect. 
At high redshifts, Ellison et al. (2001) found slightly more DLAs in the 
foreground of 
radio-selected and optically faint quasars than in optically 
selected and optically bright quasars. This result is sometimes 
misinterpreted  as evidence against dust selection effects. However, the radio-selected 
samples are still small, and the uncertainties correspondingly large. 
The difference between the true (unobscured) and observed 
column density distributions $f(N_{\rm H I})$ of DLAs 
is expected to be only $\la 50 \%$ for log $N_{\rm H I} < 21$ (see 
Fig. 1 of Pei \& Fall 1995). 
The Ellison et al. (2001) points, within the error bars, are in fact  
consistent with the predictions of 
Fall \& Pei (1993), Pei \& Fall (1995), and Pei et al. (1999). 
Recent results of Ellison et al. (2004) for Mg II systems with 
$ 0.6 < z < 1.7$ also appear to be consistent, within the large 
error bars, with the predictions of Fall \& Pei (1993), Pei \& Fall (1995), 
and Pei et al. (1999), although exact comparisons of $f(N_{\rm H I})$ 
of DLAs can only be made after 
the DLA nature of the Mg II systems is confirmed with UV spectra.
In a study of reddening in quasars with and without DLAs, 
Murphy \& Liske (2004) have recently suggested 
that the extent of dust reddening caused by high-redshift DLAs may 
be small at $z \sim 3$, but the effects at low redshift are likely to 
be larger. 
Indeed, Vladilo (2004) has reported that the dust-to-metals ratio in 
DLAs, as inferred from the depletion of Fe, increases with decreasing 
redshift. Abundance studies for a very large sample of DLAs in 
radio-selected and optically faint quasars would help to improve the 
constraints on the extent of the dust obscuration bias. 

\section{Summary}

We have obtained {\it HST} STIS UV spectra of three quasars with four 
DLAs in the redshift range $0.09 < z < 0.52$, to constrain the abundances 
of Zn, Cr, and Fe in these absorbers. These observations have provided 
the first constraints on  
Zn abundances in DLAs with $z < 0.4$. In all the 3 DLAs for which our 
observations offer meaningful constraints on the Zn abundance, 
the metallicities appear to be substantially sub-solar ($\la 10-20\%$ solar). 
We have combined our results with higher redshift data from the literature   
to estimate the global mean metallicity-redshift relation for DLAs. 
We find that the global mean metallicity of DLAs shows at most a slow increase 
with decreasing redshift and does not appear to rise up to solar or near-solar values at $z =0$.  
On the other hand, most cosmic chemical evolution models, 
based on the global star formation history inferred from the emission studies 
of galaxies, predict the global interstellar metallicity to be solar or near-solar at 
$z =0$.  
This suggests that the DLAs, especially those at low redshifts, do not 
trace the general population of galaxies 
represented by the global star formation history of the universe. 
The low 
$N({\rm H \, I})$-weighted mean metallicity of low-redshift absorbers also 
appears to contradict the fact that the mass-weighted mean interstellar metallicity of 
nearby galaxies is observed to be close to solar. 

This weak evolution in the DLA global mean metallicity could be explained 
by the fact that our sample seems to be dominated by dwarf or low surface 
brightness galaxies. It is possible that current DLA samples, especially those 
at low redshifts, are biased against more  
enriched galaxies because the latter may cause more dust obscuration of the 
background quasars. We note, however, that 
the low-redshift samples are still small, and the fact that most of the 
DLAs in our sample are metal-poor could be an effect of small 
number statistics. Metallicity measurements 
of more low-redshift DLAs are necessary to further improve the statistics of 
the mean metallicity-redshift relation and to study the scatter around the mean.
Observations of low-redshift DLAs toward optically faint quasars are 
especially necessary in the future to 
improve the constraints on the mean metallicity-redshift relation for 
DLAs, and to quantify potential dust selection effects. 

\acknowledgments

VPK, SMF, and JTL acknowledge partial 
support from the NASA / Space Telescope Science Institute grant GO-9441. 
VPK also acknowledges partial support from the 
National Science Foundation grant AST-0206197. DEW acknowledges 
support from the NASA LTSA grant NAG5-11413. JWT acknowledges support of 
the NSF Physics Frontier Center, Joint Institute for Nuclear Astrophysics 
under grant PHY 02-16783 and DOE support under grant DE-FG 02-91ER 40606.
We thank Drs. Steven Beckwith, Bahram Mobasher, Charles Proffitt 
and Anthony Roman of STScI for help with execution 
and calibration of our STIS observations. We also thank  
J. Meiring for assistance with data analysis. Finally, we thank an 
anonymous referee for helpful comments on the paper.  



Facilities: \facility{HST(STIS)}.




\clearpage

\begin{figure}
\epsscale{1.0}
\plotone{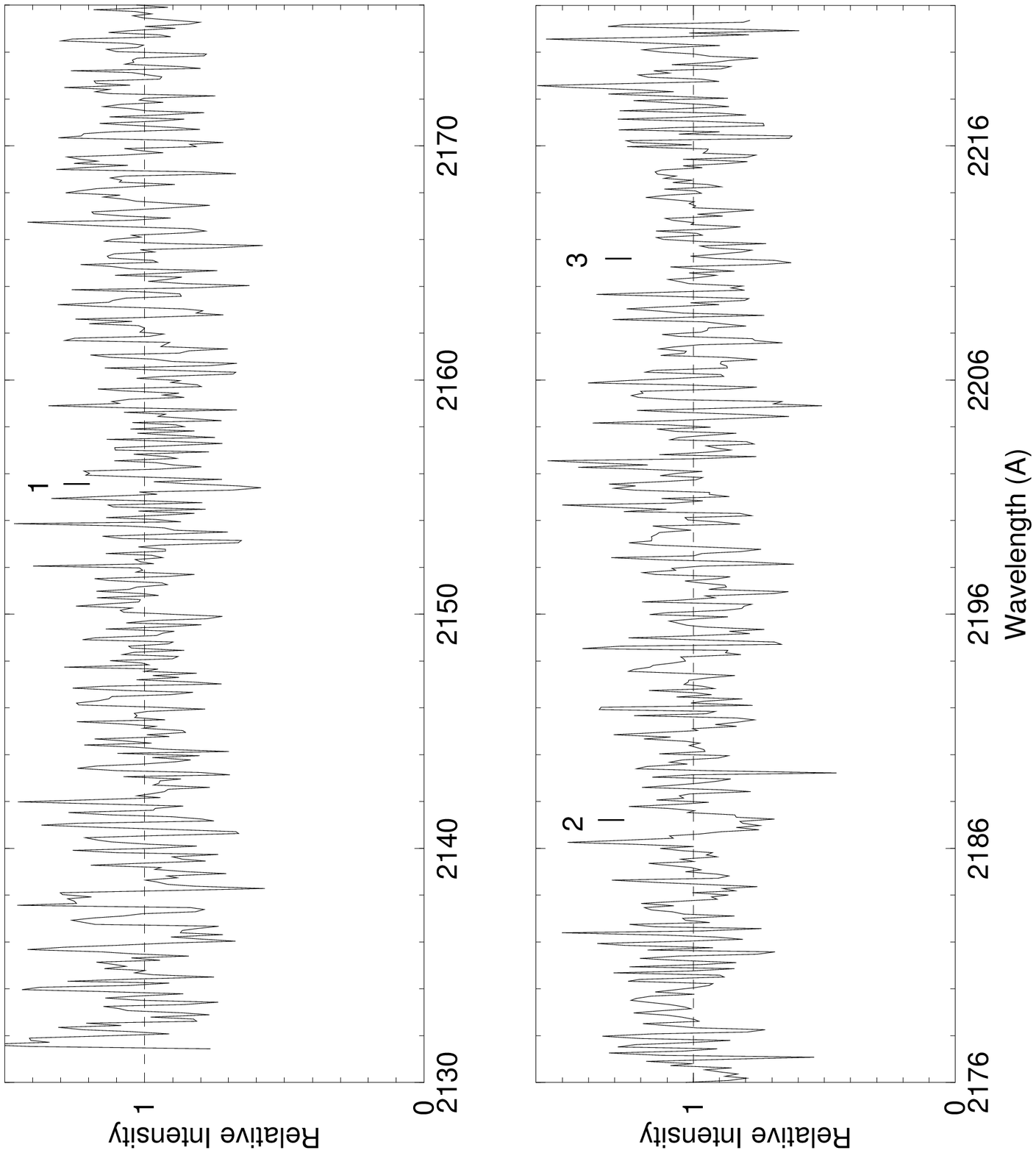}
\caption{HST STIS spectrum of Q0738+313 observed with the G230M grating
setting centered at 2176 {\AA}. \label{fig1}}
\end{figure}
\clearpage

\begin{figure}
\epsscale{1.0}
\plotone{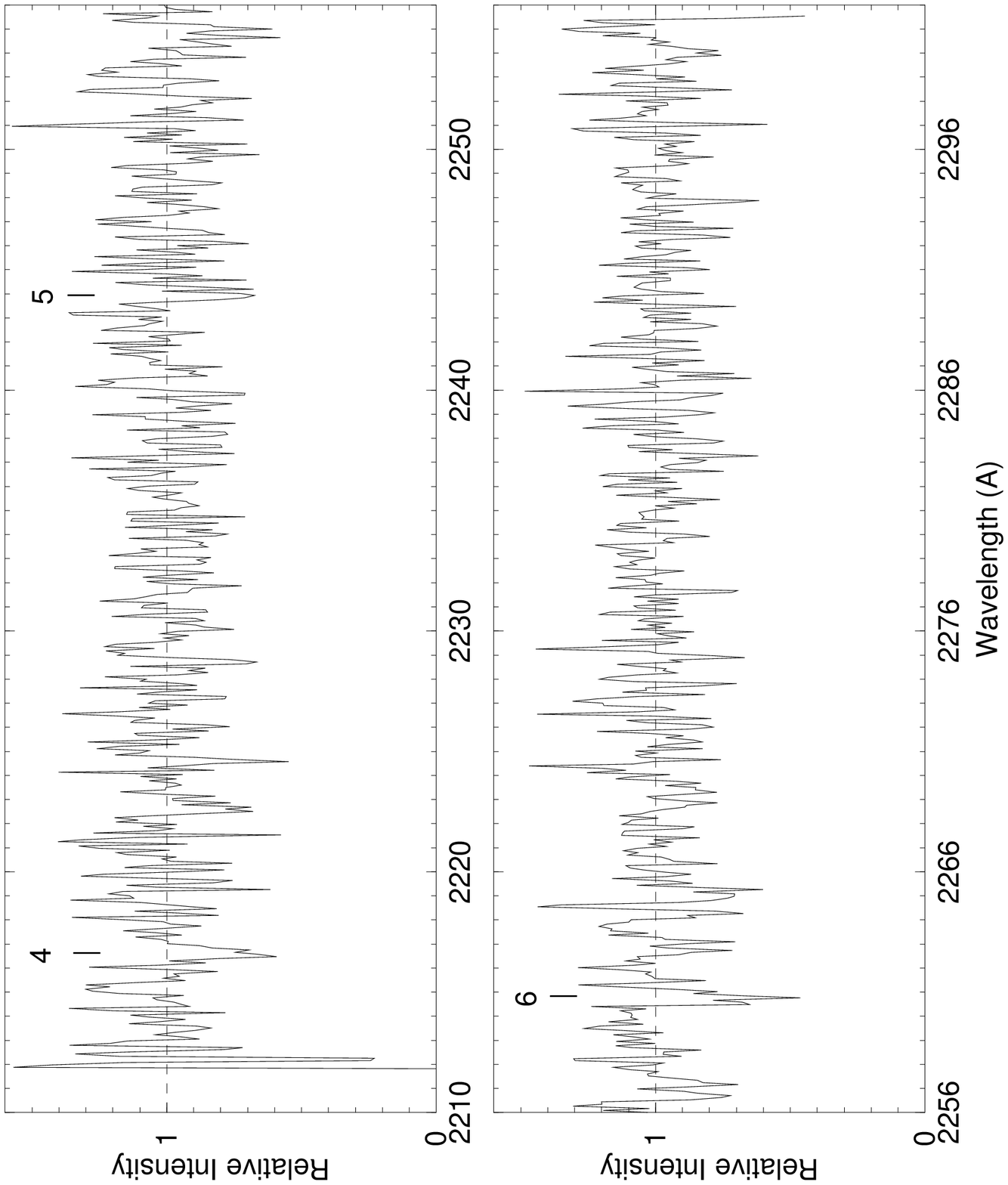}
\caption{HST STIS spectrum of Q0738+313 observed with the  G230M grating 
setting centered at 2257 {\AA}. \label{fig2}}
\end{figure}
\clearpage

\begin{figure}
\epsscale{0.75}
\plotone{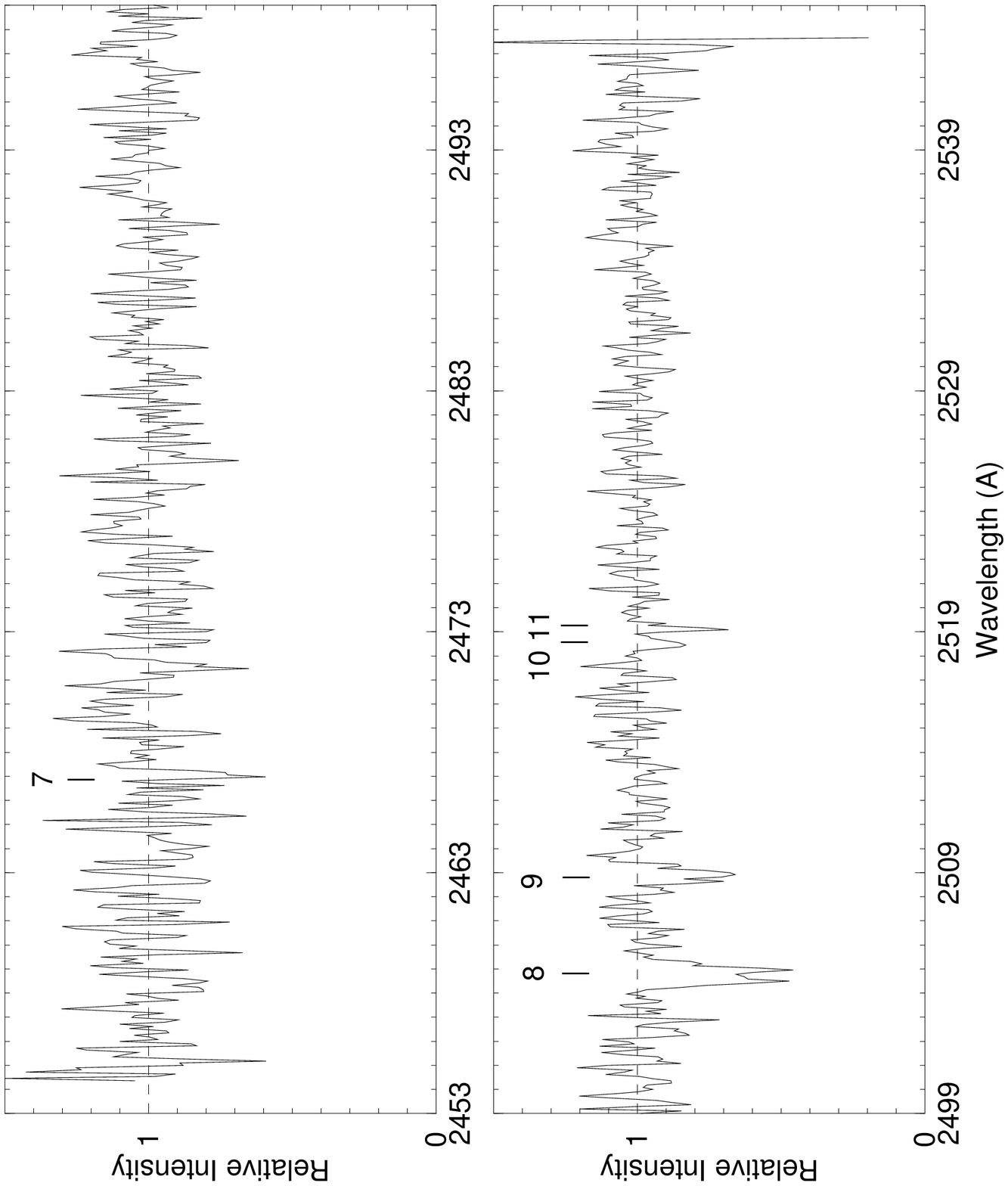}
\caption{HST STIS spectrum of Q0738+313 observed with the G230M grating 
setting centered at 2499 {\AA}. \label{fig3}}
\end{figure}
\clearpage

\begin{figure}
\epsscale{1.0}
\plotone{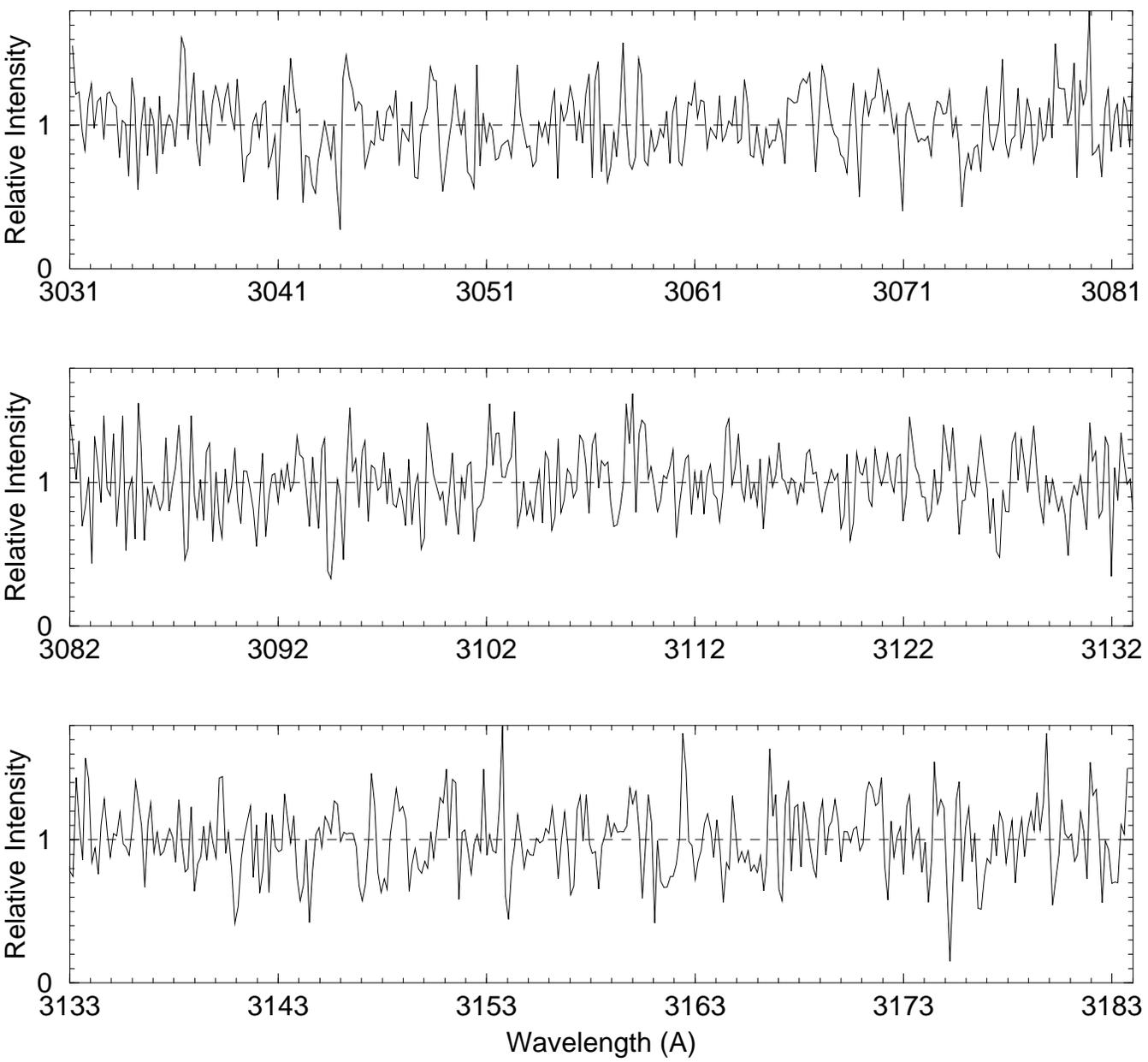}
\caption{HST STIS spectrum of Q0827+243 observed with the G230MB grating  
setting centered at 3115 {\AA}. \label{fig4}}
\end{figure}
\clearpage

\begin{figure}
\epsscale{0.7}
\plotone{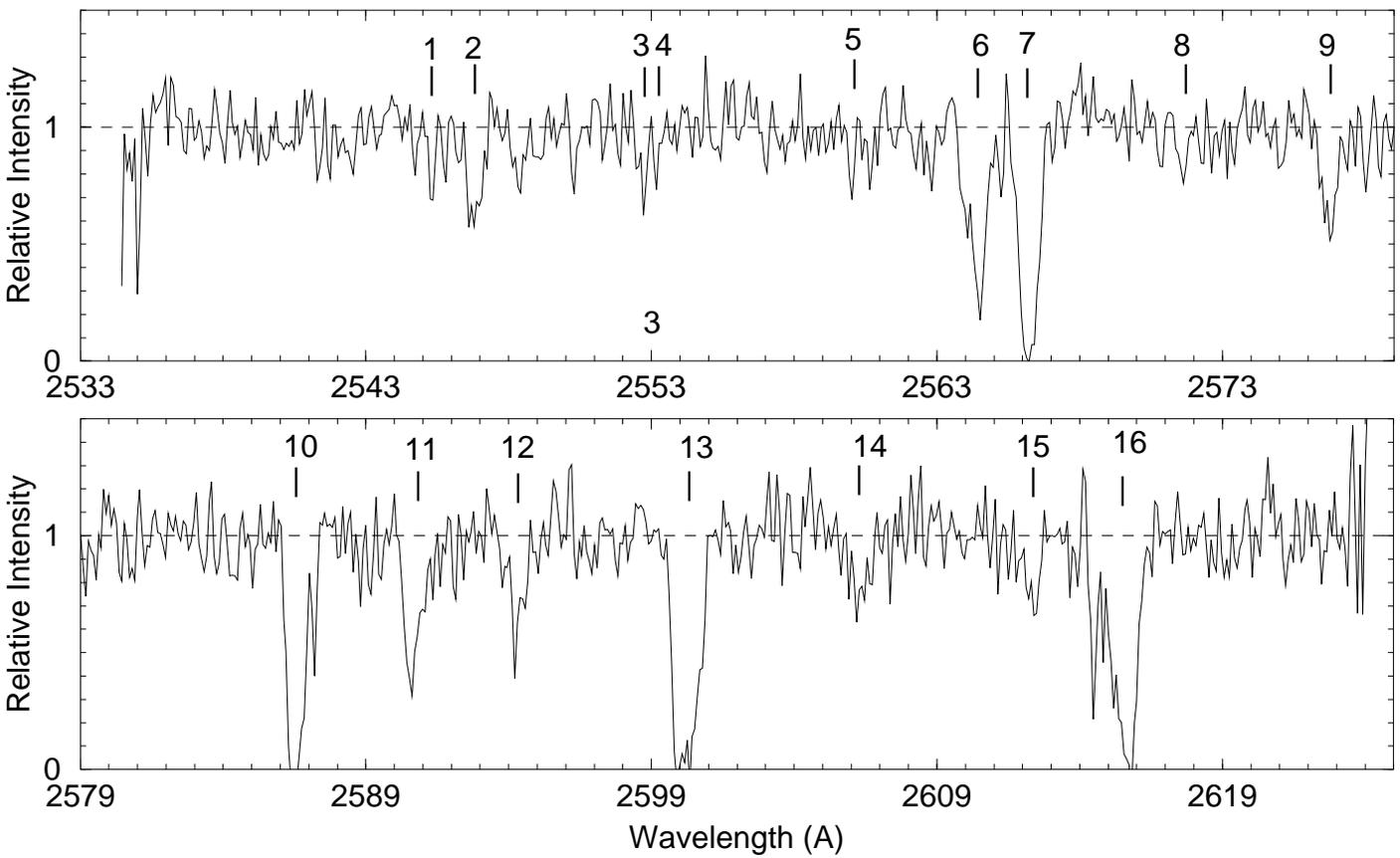}
\caption{HST STIS spectrum of Q0952+179 observed with the G230M grating 
setting centered at 2579 {\AA}. \label{fig5}}
\end{figure}
\clearpage

\begin{figure}
\epsscale{1.0}
\plotone{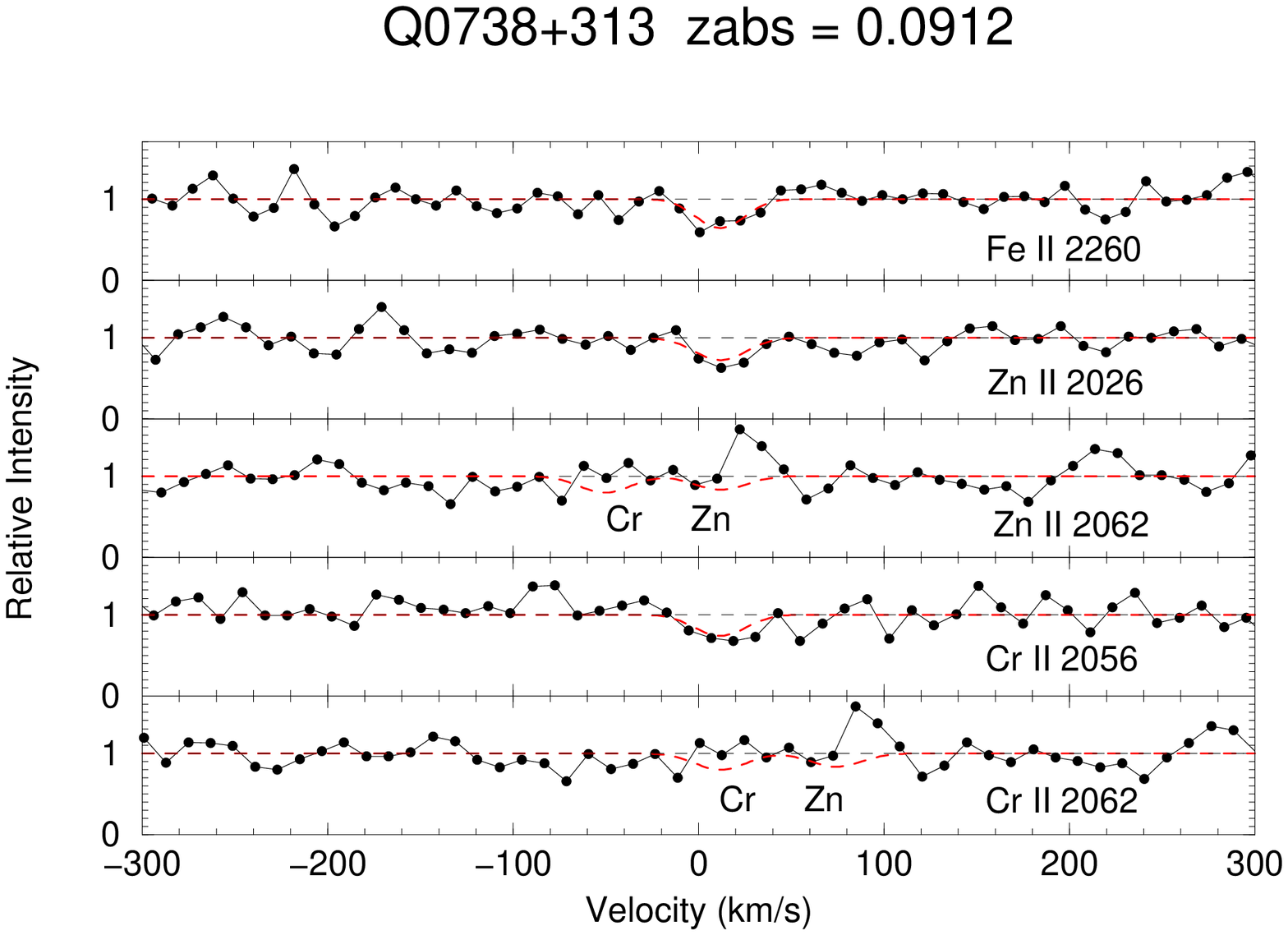}
\caption{Velocity plot of the Fe II, Zn II, and Cr II data for 
the $z=0.0912$ DLA toward Q0738+313. The zero 
point of the velocity scale 
refers to the absorption redshift $z=0.0912$ inferred from the 21-cm and 
optical data. The dashed curves show the profiles for the best-fit 
$N$, $b$, and $v$ parameters obtained by fitting multiple lines 
simultaneously. In the case of Zn II, the profiles correspond 
to the 3 $\sigma$ upper limit adopted. \label{fig6}}
\end{figure}
\clearpage
 
\begin{figure}
\epsscale{0.8}
\plotone{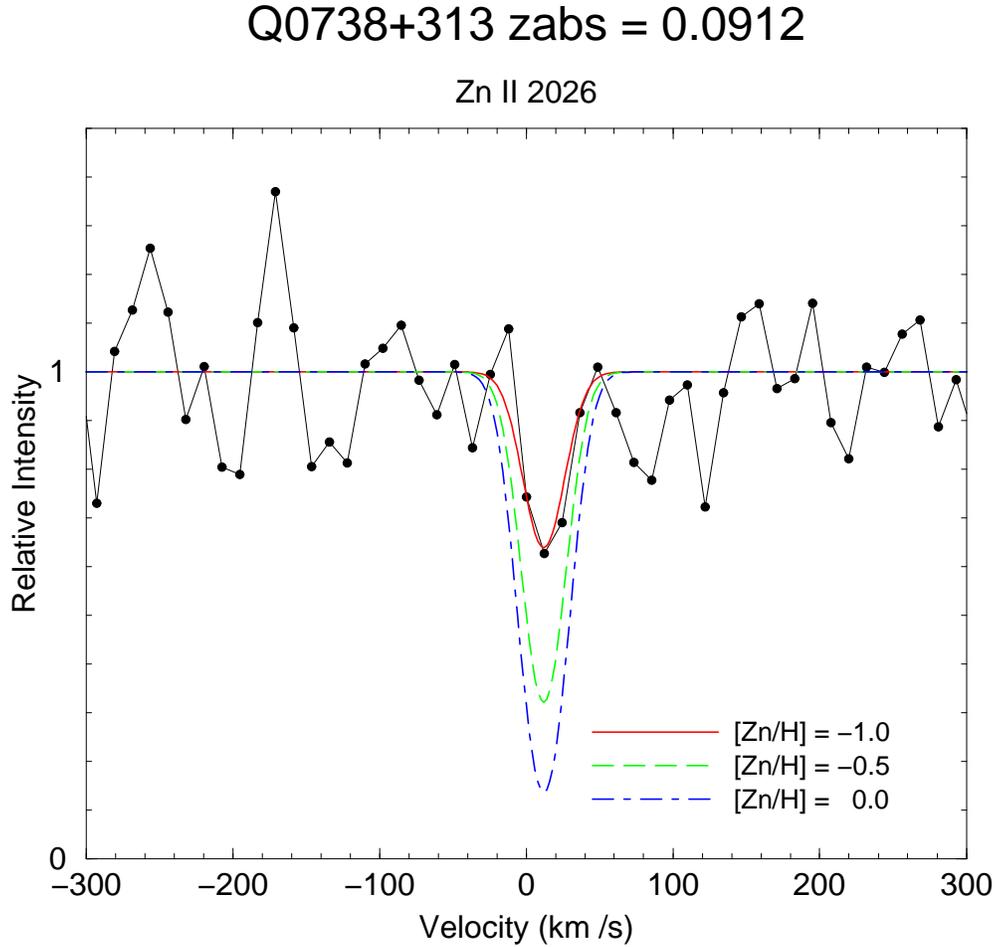}
\caption{Comparison of Zn II $\lambda 2026$ data for the $z=0.0912$ DLA 
toward Q0738+313  
with simulated Voigt profiles for [Zn/H] = -1.0, -0.5, and 0.0, shown by 
solid (red), short-dashed (green), and short-dash-long-dashed (blue) 
curves, respectively. All 
three curves assume the best-fit $b$ and $v$ values 
obtained by fitting multiple lines 
simultaneously. \label{fig7}} 
\end{figure} 
\clearpage

\begin{figure}
\epsscale{0.8}
\plotone{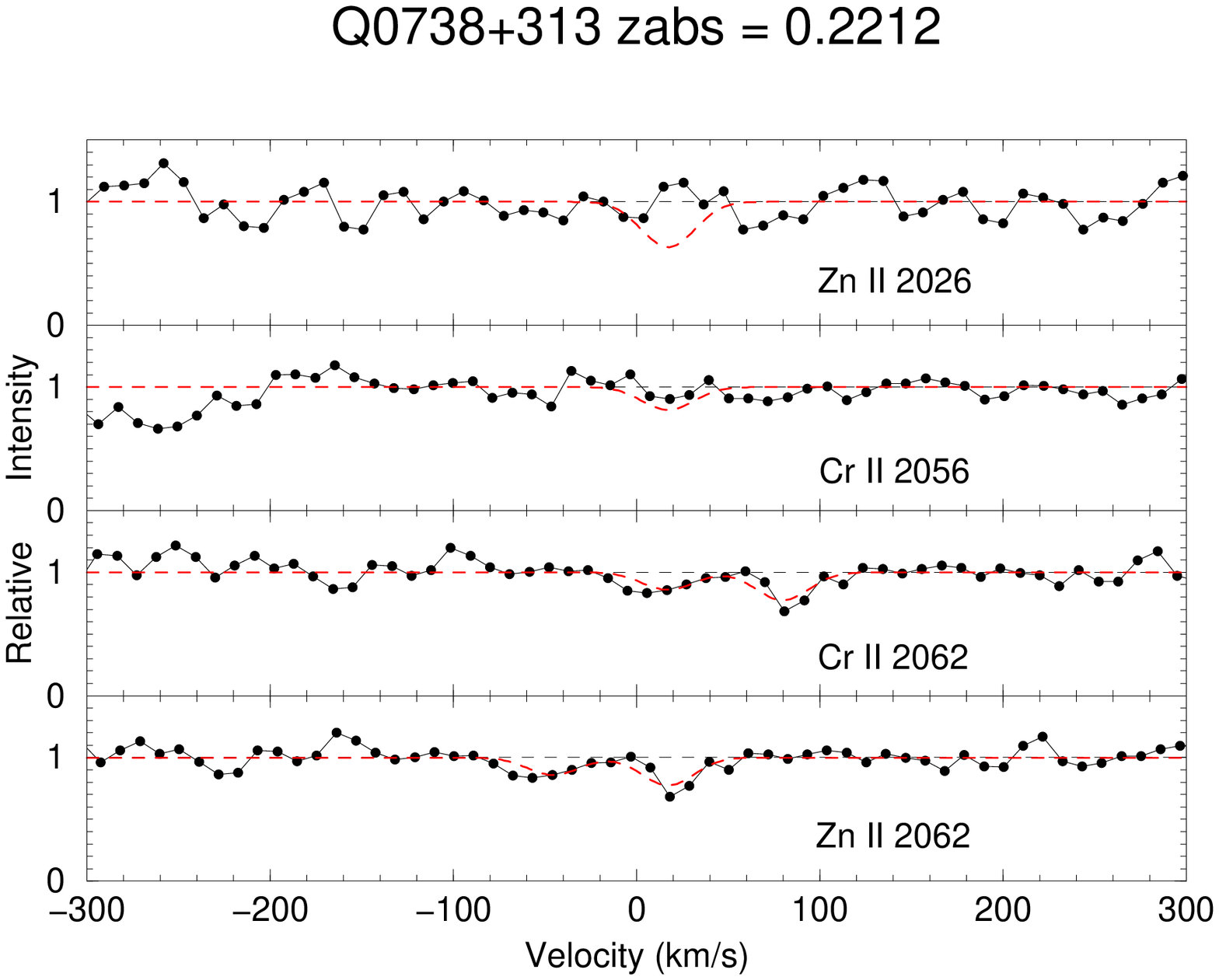}
\caption{Velocity plot of the Zn II and Cr II data for 
the $z=0.2212$ DLA toward Q0738+313. The zero 
point of the velocity scale 
refers to the absorption redshift $z=0.2212$ inferred from the 21-cm and 
optical data. The dashed curves show the profiles for the best-fit 
$N$, $b$, and $v$ parameters obtained by fitting multiple lines 
simultaneously. In the case of Zn II, the profiles correspond 
to the 3 $\sigma$ upper limit adopted.\label{fig8}}
\end{figure}
\clearpage

\begin{figure}
\epsscale{0.8}
\plotone{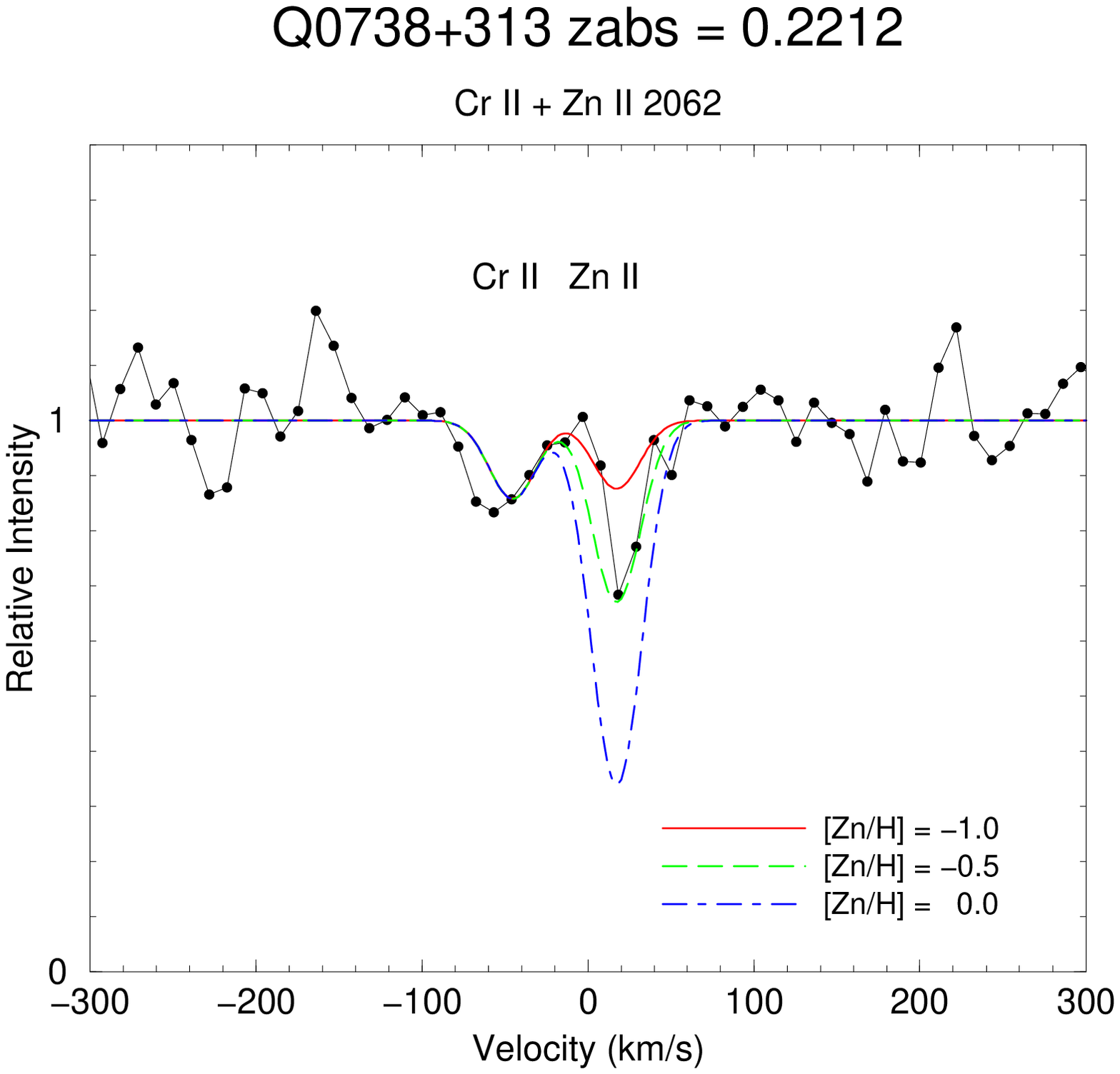}
\caption{Comparison of Zn II $\lambda 2062.664$ data for the $z=0.2212$ DLA 
toward Q0738+313 with simulated Voigt profiles for 
[Zn/H] = -1.0, -0.5, and 0.0, shown by 
solid (red), short-dashed (green), and short-dash-long-dashed (blue) curves, respectively. 
All three curves 
assume the best-fit values of $b$, $v$, and Cr II column density 
obtained by fitting multiple lines 
simultaneously. \label{fig9}} 
\end{figure} 
\clearpage

 \begin{figure}
\epsscale{0.8}
 \plotone{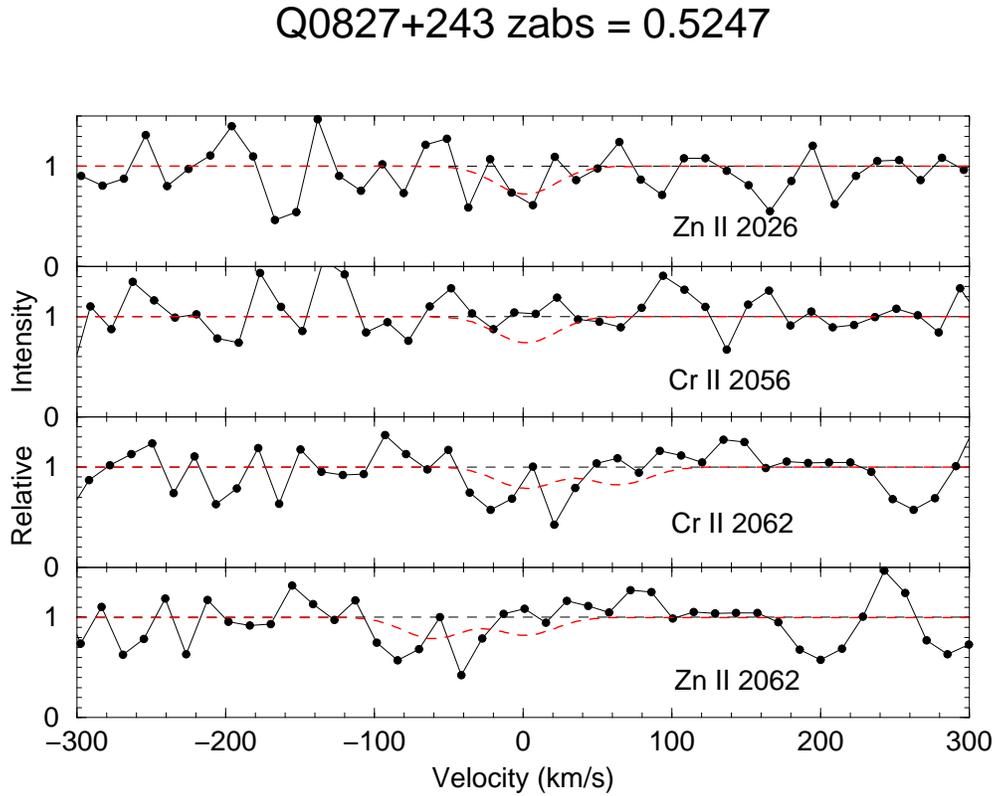}
 \caption{Velocity plot of the Zn II and Cr II data for 
 the $z=0.5247$ DLA toward Q0827+243. The zero 
point of the velocity scale 
refers to the absorption redshift $z=0.5247$ inferred from the 21-cm and 
optical data. The dashed curves show profiles corresponding  
to the 3 $\sigma$ upper limits adopted for $b = 10$ km s$^{-1}$. 
\label{fig10}}
 \end{figure}
\clearpage

\begin{figure}
\epsscale{0.8}
\plotone{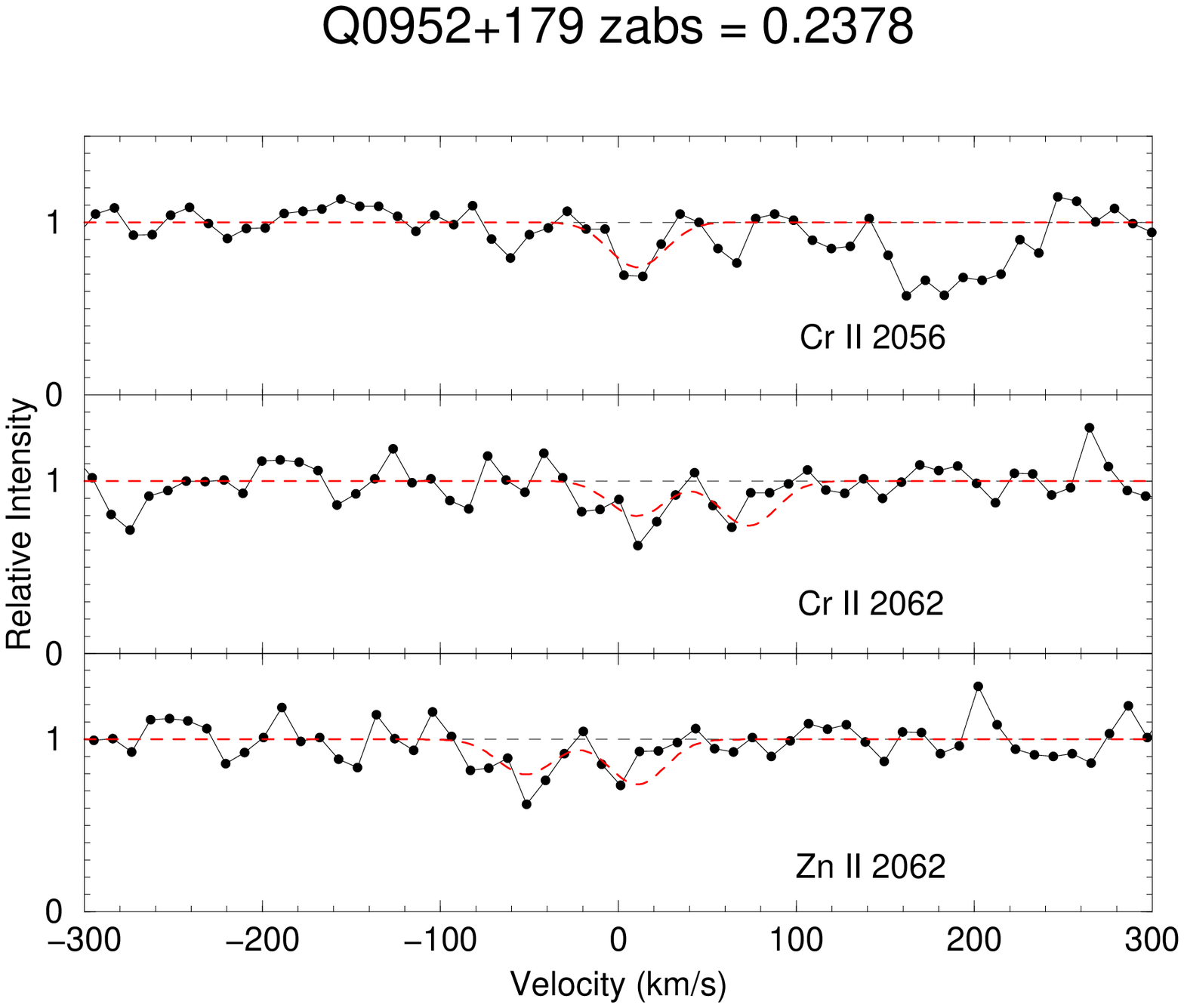}
\caption{Velocity plot of the Zn II and Cr II data for 
the $z=0.2378$ DLA toward Q0952+179. The zero 
point of the velocity scale 
refers to the absorption redshift $z=0.2378$ inferred from the 21-cm and 
optical data. The dashed curves show the profiles for the best-fit 
$N$, $b$, and $v$ parameters obtained by fitting multiple lines 
simultaneously. In the case of Zn II, the profiles correspond 
to the 3 $\sigma$ upper limit adopted. \label{fig11}}
\end{figure}

\clearpage
\begin{figure}
\epsscale{0.8}
\plotone{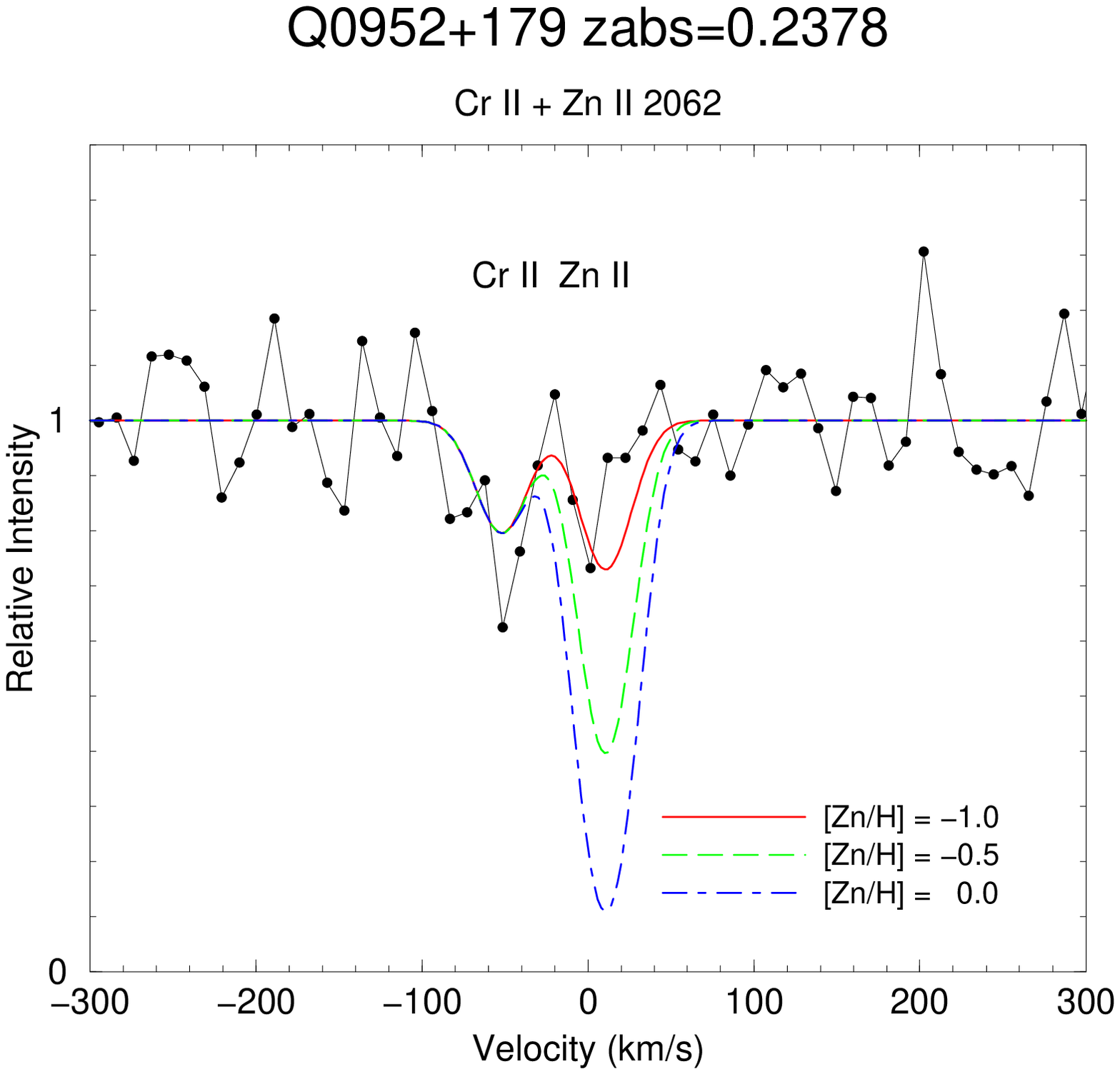} 
\caption{Comparison of Zn II $\lambda 2062$ data for the $z=0.2378$ DLA 
toward Q0952+179 with simulated Voigt profiles for [Zn/H] = -1.0, -0.5, 
and 0.0, shown by solid (red), short-dashed (green), and 
short-dash-long-dashed curves (blue), respectively. All three curves 
assume the best-fit values of $b$, $v$, and Cr II column density 
obtained by fitting multiple lines simultaneously.  \label{fig12}} 
\end{figure}
\clearpage 

\begin{figure}
\epsscale{1.0}
\plotone{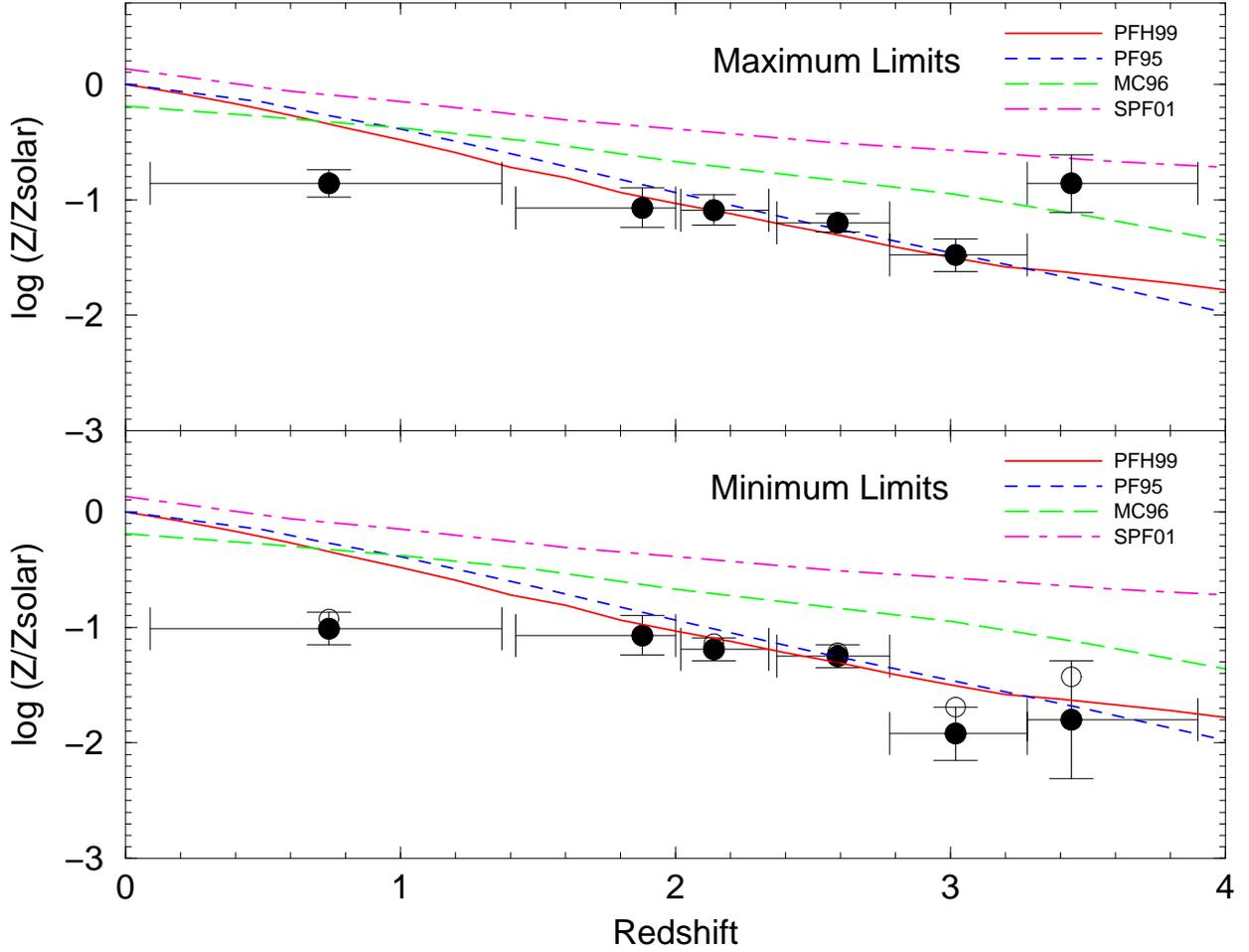} 
\caption{The global metallicity-redshift relation deduced from 
damped Ly-$\alpha$ absorbers. 
The circles show the logarithm of the  
$N({\rm H \, I})$-weighted mean Zn metallicity relative to the solar value  
vs. redshift for the sample of 87 DLAs based on our {\it HST} observations 
and the literature. The filled circles in the top (a) and bottom (b) panels 
show, respectively, the $N({\rm H \, I})$-weighted mean metallicity for 
the maximum-limits and minimum-limits cases using Zn only.  The unfilled 
circles in the bottom panel show the $N({\rm H \, I})$-weighted mean 
metallicity for the  
modified minimum-limits sample, using other elements to constrain 
the metallicities in cases of Zn limits.  
Vertical error bars on the filled circles in both panels denote 1 $\sigma$ uncertainties 
in the logarithm of the $N({\rm H \, I})$-weighted mean metallicity. 
Data points are plotted at the median redshift in each bin. Horizontal 
bars for each bin extend from the lowest DLA redshift to the highest 
DLA redshift within that bin. The short-dashed (blue), long-dashed 
(green), solid (red), and short-dash-long-dashed (magenta) curves show, 
respectively, the ``true'' mean metallicity (not corrected for dust 
obscuration) expected in the cosmic chemical evolution models 
of Pei \& Fall (1995),  Malaney \& Chaboyer (1996), Pei et al. (1999), 
and Somerville et al. (2001). \label{fig13}} 
\end{figure}






\clearpage

\begin{table}
\centerline{\bf{ TABLE 1}}
\centerline{\bf{ Target DLAs}}
\vskip15pt
\begin{center}
\begin{tabular}{|l|l|l|l|l|}
\tableline
QSO & $z_{em}$ & $z_{abs}$& log $N_{\rm H I}$& Ref.$^{\dagger}$\\
\tableline
\tableline
Q0738+313&0.635&0.0912&$21.18^{+0.05}_{-0.06}$&1, 2\\
Q0738+313&0.635&0.2212&$20.90^{+0.07}_{-0.08}$&1, 3\\
Q0827+243&0.939&0.5247&$20.30^{+0.04}_{-0.05}$&4, 5\\
Q0952+179&1.472&0.2378&$21.32^{+0.05}_{-0.06}$&4, 5\\
\tableline
\end{tabular}
\end{center}
\noindent{ $\dagger$ References: 1. Rao \& Turnshek (1998); 
2. Chengalur \& Kanekar (1999); 
3. Lane et al. (1998); 4. Rao \& Turnshek (2000); 5. Kanekar \& Chengalur (2001).}
\end{table}
\clearpage

\begin{table}
\centerline{\bf{ TABLE 2}}
\centerline{\bf{Details of {\it HST} STIS Observations }}
\vskip15pt
\begin{center}
\begin{tabular}{|l|l|l|l|l|l|}
\tableline
QSO& Detector&Grating& $R$&$\lambda_{\rm{cent}}$&$t_{\rm exp}$ (s)\\
\tableline
\tableline
Q0738+313&NUV-MAMA&G230M&12,100&2176& 13,388\\
Q0738+313& NUV-MAMA&G230M&12,500& 2257& 11,224\\
Q0738+313&NUV-MAMA&G230M&13,900&2499& 13,613\\
Q0827+243&CCD&G230MB&10,400&3115&12,232\\
Q0952+179&NUV-MAMA&G230M&14,300&2579&24,346\\
\tableline
\end{tabular}
\end{center}
\end{table}
\clearpage

\begin{table}
\centerline{\bf {TABLE 3}}
\centerline{\bf{Line Lists}}
\bigskip
\begin{center}
\begin{tabular}{|l|l|l|l|l|l|l|}
\tableline
QSO&No.&$\lambda_{obs}$&$W_{obs}$&$\sigma_{W}^{p}$&ID&$z_{abs}$\\
&&({\AA})&({\AA})&({\AA})&&\\
\tableline
\tableline
Q0738+313&1$^{\dagger}$&2155.47 & 0.140 & 0.039&&\\
&2$^{\dagger}$&2187.08 & 0.157 & 0.043&&\\
&3&2211.03 & 0.092 & 0.027& Zn II 2026& 0.0912\\
&4$^{\dagger}$&2216.58 & 0.157 & 0.039&&\\
&5&2243.99 & 0.096 & 0.029&Cr II 2056 & 0.0912\\
&6&2260.75 & 0.180 & 0.033&Fe II 2260 & 0.0000\\
&7&2467.07 & 0.109 &0.028& Fe II 2260 & 0.0912\\
&8&2504.76 & 0.403 & 0.036& C IV 1548 & 0.6179\\
&9&2508.86 & 0.234& 0.030& C IV 1551& 0.6178\\
&10&2518.49 & 0.061& 0.020&Cr II 2062& 0.2212\\
&11&2519.13 & 0.066& 0.017&Zn II 2062 & 0.2213\\
\tableline
Q0952+179&1&2545.32& 0.070 & 0.020 & Cr II 2056 & 0.2378\\
&2&2546.78 &0.234&0.032 & Ly-$\alpha$& 1.0950\\
&3&2552.69& 0.100 & 0.024& Cr II 2062& 0.2378\\
&4&2553.18&0.052&0.021&Zn II 2062&0.2378\\
&5&2560.01&0.060&0.019&Ly-$\alpha$&1.1058\\
&6&2564.45&0.616&0.037&Ly-$\alpha$&1.1095\\
&7&2566.22&0.770&0.033&Ly-$\alpha$&1.1109\\
&8&2571.61&0.081&0.025&Ly-$\alpha$&1.1154\\
&9&2576.78&0.282&0.036&Mn II 2577& 0.0003\\
&10&2586.60&0.843&0.042&Fe II 2586& 0.0003\\
&11&2590.69&0.447&0.044&Ly-$\alpha$&1.1311\\
&12&2594.27&0.262&0.039&Mn II 2594&0.0002\\
&13&2600.21&1.054&0.043&FeII 2600&0.0003\\
&14&2606.26&0.218&0.040&Mn II 2606&0.0002\\
&15&2612.30&0.198&0.029&Ly-$\alpha$&1.1489\\
&16&2615.42&1.224&0.057&Ly-$\alpha$&1.1514\\
\tableline
\end{tabular}
\end{center}
\hskip40pt {$\dagger$ No clear identification, but possibly 
C IV $\lambda$ 1548 lines at $z_{abs} = 0.392$, 0.413, and 0.432 with 
the corresponding C IV $\lambda 1551$ lines lost in noise.} 
\end{table}
\clearpage

\begin{table}
\centerline{\bf {TABLE 4}}
\centerline{\bf{Further Details on DLA Metal-line Features}}
\bigskip
\begin{center}
\begin{tabular}{|l|l|l|l|l|l|l|l|}
\tableline
QSO&$\lambda_{obs}$&$W_{obs}$&$\sigma_{W}^{p}$&$\sigma_{W}^{c}$&
$\sigma_{W}^{t}$&ID&$z_{abs}$\\
&({\AA})&({\AA})&({\AA})&({\AA})&({\AA})&&\\
\tableline
\tableline
Q0738+313&2211.03 & 0.092 & 0.027& 0.019& 0.033& Zn II 2026& 0.0912\\
&2243.99 & 0.096 & 0.029& 0.020 & 0.036& Cr II 2056 & 0.0912\\
&2467.07 & 0.109 &0.028& 0.018 & 0.033& Fe II 2260 & 0.0912\\
\tableline
Q0738+313&2518.49 & 0.061& 0.020&0.009& 0.022&Cr II 2062& 0.2212\\
&2519.13 & 0.066& 0.017& 0.008& 0.019& Zn II 2062 & 0.2213\\
\tableline
Q0952+179&2545.32& 0.070 & 0.020 &0.019&0.028& Cr II 2056 & 0.2378\\
&2552.69& 0.100 & 0.021 & 0.012 & 0.024& Cr II 2062& 0.2378\\
&2553.18&0.052&0.021&0.013&0.025& Zn II 2062&0.2378\\
\tableline
\end{tabular}
\end{center}
\end{table}
\clearpage
\begin{table}
\centerline{\bf {TABLE 5}}
\centerline{\bf{Column Densities and Abundances for the 
$z = 0.0912$ DLA toward Q0738+313}}
\vskip15pt
\begin{center}
\begin {tabular}{|c|c|c|c|c|c|c|c|}
\tableline
Fitting&$b$ & log $N_{\rm Cr II}$ & log $N_{\rm Zn II}$& 
log $N_{\rm Fe II}$&[Cr/H] & 
[Zn/H]&[Fe/H]\\
Procedure&(km&(cm$^{-2}$)&(cm$^{-2}$)&(cm$^{-2}$)&&&\\
&s$^{-1}$)&&&&&&\\
\tableline
\tableline
$N$,$b$,$v$ varied$^{\dagger}$;&
$11.2$&$13.28^{+0.14}_{-0.22}$ &$< 12.66$&$15.02^{+0.12}_{-0.17}$ 
&$-1.55^{+0.15}_{-0.23}$ &$< -1.14$& $-1.63^{+0.13}_{-0.18}$\\
multiple lines&$\pm 6.7$&&&&&&\\
&&&&&&&\\
$b$ fixed;&
5.0 & $13.56^{+0.24}_{-0.56}$&$13.06^{+0.22}_{-0.49}$&$15.22^{+0.21}_{-0.41}$&
$-1.27^{+0.25}_{-0.56}$&$-0.75^{+0.23}_{-0.49}$&$-1.43^{+0.22}_{-0.41}$\\
single lines&10.0& $13.49^{+0.16}_{-0.24}$&$12.84^{+0.13}_{-0.19}$&$15.06^{+0.13}_{-0.20}$&
$-1.34^{+0.17}_{-0.25}$&$-0.97^{+0.17}_{-0.20}$&$-1.59^{+0.17}_{-0.21}$\\
&15.0& $13.48^{+0.14}_{-0.21}$&$12.81^{+0.12}_{-0.17}$&$15.06^{+0.12}_{-0.17}$&
$-1.35^{+0.15}_{-0.22}$&$-1.00^{+0.13}_{-0.18}$&$-1.59^{+0.13}_{-0.18}$\\
&20.0& $13.50^{+0.13}_{-0.19}$&$12.81^{+0.12}_{-0.17}$&$15.06^{+0.12}_{-0.16}$&
$-1.33^{+0.17}_{-0.20}$&$-1.00^{+0.13}_{-0.18}$&$-1.59^{+0.13}_{-0.17}$\\

\tableline
\end{tabular}
\end{center}
\hskip40pt {$\dagger$ For the fits with $N, b, v$ varied, the error bars on column densities and abundances 
include uncertainties 
from Poisson statistics and uncertainties in $b$ value, 
central velocity, and continuum placement. The limits are $3 \, \sigma$ 
upper limits.}
\end{table}
\clearpage

\begin{table}
\centerline{\bf {TABLE 6}}
\centerline{\bf{Column Densities and Abundances for the 
$z = 0.2212$ DLA toward Q0738+313}}
\vskip15pt
\begin{center}
\begin {tabular}{|c|c|c|c|c|c|}
\tableline
Fitting Procedure&$b$ & log $N_{\rm Cr II}$ & log $N_{\rm Zn II}$& [Cr/H] & 
[Zn/H]\\
&(km s$^{-1}$)&(cm$^{-2}$)&(cm$^{-2}$)&&\\
\tableline
\tableline
$N$,$b$,$v$ varied$^{\dagger}$; multiple lines$^{\dagger \dagger}$&
$12.2 \pm 7.5$&$13.11^{+0.15}_{-0.24}$&
$12.83^{+0.12}_{-0.16}$&$-1.44^{+0.17}_{-0.25}$& $-0.70^{+0.14}_{-0.17}$ \\
&&&&&\\
$b$ fixed; single lines&
5.0 & $13.22^{+0.16}_{-0.25}$& $12.75^{+0.15}_{-0.24}$&
$-1.33^{+0.18}_{-0.26}$& $-0.78^{+0.17}_{-0.25}$\\
&10.0&$13.20^{+0.14}_{-0.20}$ & $12.73^{+0.13}_{-0.18}$& 
$-1.35^{+0.16}_{-0.21}$& $-0.80^{+0.15}_{-0.19}$\\
&15.0&$13.21^{+0.13}_{-0.19}$&$12.76^{+0.12}_{-0.17}$& 
$-1.34^{+0.15}_{-0.20}$ &$-0.77^{+0.14}_{-0.18}$\\
&20.0&$13.23^{+0.13}_{-0.19}$&$12.80^{+0.12}_{-0.16}$& 
$-1.32^{+0.15}_{-0.20}$ &$-0.73^{+0.14}_{-0.17}$\\
\tableline
\end{tabular}
\end{center}
\hskip40pt {$\dagger$: For the fits with $N, b, v$ varied, the error 
bars on column densities and abundances include uncertainties 
from Poisson statistics and uncertainties in $b$ value, 
central velocity, and continuum placement.
$\dagger \dagger$: The noisy data at the positions of the $\lambda 2026$ 
and $\lambda 2056$ lines are not included 
in these fits, but they  
give consistent results of log $N_{\rm Zn II} < 12.81$ and log $N_{\rm Cr II} < 13.37$, 
respectively for the best-fit $b$ value of 12.2 km s$^{-1}$. 
Nevertheless, to be conservative, we finally adopt an upper 
limit [Zn/H]$ \le -0.70$ for this system.
}
\end{table}

\clearpage
\begin{table}
\centerline{\bf {TABLE 7}}
\centerline{\bf{Column Densities 
and Abundances for the $z = 0.5247$ DLA 
toward Q0827+243}}
\vskip15pt
\begin{center}
\begin {tabular}{|c|c|c|c|c|c|}
\tableline
Fitting Procedure&$b$ & log $N_{\rm Cr II}$ & log $N_{\rm Zn II}$& [Cr/H] & 
[Zn/H]\\
&(km s$^{-1}$)&(cm$^{-2}$)&(cm$^{-2}$)&&\\
\tableline
\tableline
$b$ fixed; single lines$^{\dagger}$ &5.0 & $< 13.79$ &$< 13.23$ &$< -0.16$&  $<  0.30$ \\
&10.0& $< 13.51$ &$< 12.89$ &$< -0.44$&  $< -0.04$ \\
&15.0& $< 13.45$ &$< 12.83$ &$< -0.50$&  $< -0.10$ \\
&20.0& $< 13.42$ &$< 12.80$ &$< -0.53$&  $< -0.13$ \\
\tableline
\end{tabular}
\end{center}
\hskip40pt {$\dagger$ The limits are $3 \, \sigma$ upper limits.}
\end{table}
\clearpage

\begin{table}
\centerline{\bf {TABLE 8}}
\centerline{\bf{Column Densities and Abundances for the 
$z = 0.2378$ DLA toward Q0952+179}}
\vskip15pt
\begin{center}
\begin {tabular}{|c|c|c|c|c|c|}
\tableline
Fitting Procedure&$b$ & log $N_{\rm Cr II}$ & log $N_{\rm Zn II}$& [Cr/H] & 
[Zn/H]\\
&(km s$^{-1}$)&(cm$^{-2}$)&(cm$^{-2}$)&&\\
\tableline
\tableline$N$,$b$,$v$ varied$^{\dagger}$;&
$15.0 \pm 7.7$&$13.32^{+0.12}_{-0.17}$ &$< 12.93$& $-1.65^{+0.14}_{-0.18}$ &$< -1.02$ \\
multiple lines&&&&&\\
&&&&&\\
$b$ fixed; single lines&
5.0 & $13.30 \pm 0.20$ & $12.62\pm 0.20$& $-1.67 \pm 0.21$ 
& $-1.33 \pm 0.25$\\
&10.0& $13.24 \pm 0.20$ & $12.62\pm 0.20$& $-1.73 \pm 0.21$ 
& $-1.33 \pm 0.25$\\
&15.0& $13.25 \pm 0.20$ & $12.66\pm 0.20$& $-1.72 \pm 0.21$ 
& $-1.29 \pm 0.25$\\
&20.0& $13.26 \pm 0.20$& $12.71\pm 0.20$& $-1.71 \pm 0.21$  
& $-1.24 \pm 0.25$\\

\tableline
\end{tabular}
\end{center}
\hskip40pt {$\dagger$: For the fits with $N, b, v$ varied, the error bars on column densities and abundances include uncertainties 
from Poisson statistics and uncertainties in $b$ value, 
central velocity, and continuum placement. 
The limits are $3 \, \sigma$ upper limits.}
\end{table}
\clearpage

\begin{table}
\centerline{\bf {TABLE 9}}
\centerline{\bf{Summary of Abundances}}
\vskip15pt
\begin{center}
\begin {tabular}{|l|l|l|l|l|l|}
\tableline
QSO&$z_{abs}$&[Zn/H]$^{*}$&[Cr/H]$^{*}$&[Fe/H]& Ref.$^{\dagger}$ \\
\tableline
\tableline
Q0738+313&0.0912&$< -1.14$&$-1.55^{+0.15}_{-0.23}$& 
$-1.63^{+0.13}_{-0.18}$&1\\
Q0738+313&0.2212&$\le -0.70$&$-1.44^{+0.17}_{-0.25}$&...&1\\
Q0827+243&0.5247&$< -0.04$&$< -0.44$&$-1.02 \pm 0.05$ &1, 2\\
Q0952+179&0.2378&$< -1.02$&$-1.65^{+0.14}_{-0.18}$&...&1\\
\tableline
\end{tabular}
\end{center}
\hskip35pt { $*$: Values obtained from fits with $N$, $b$, $v$ varied 
simultaneously for multiple lines.}
\hskip40pt {$\dagger$ References: 1. This Work; 2. Khare et al. (2004).} 
\end{table}
\clearpage

\begin{table}
\centerline{\bf {TABLE 10}}
\centerline{\bf{Summary of Zn Measurements in DLAs}}
\vskip15pt
\begin{center}
\begin {tabular}{|l|c|l|l|l|l|}
\tableline
QSO & Ref.$^{\dagger}$&  $z_{abs}$& [Zn/H]&$N_{\rm H I}/10^{20}$(cm$^{-2}$)& $N_{\rm Zn II}/10^{12}$(cm$^{-2}$)\\
\tableline
\tableline
0738+313&  1& 	0.0912&	$\le  -1.14$&	$15.00  \pm 2.00$&	$\le 4.59$\\
0738+313&  1&	0.2212&	$\le -0.70$&	$7.90	\pm 1.40$&	$\le 6.69$\\
0952+179&  1&	0.2378&	$\le  -1.02$&	$21.00	\pm 2.50$&	$\le 8.55$\\
1229-021&  2&	0.3950&	-0.45&	$5.62	\pm 0.91$&	$8.51	\pm 2.38$\\
0827+243&  1&	0.5247&	$\le  -0.04$&	$2.00	\pm 0.20$&	$\le 7.75$\\
1122-1649& 3&	0.6819&	$\le  -1.29$&	$2.82	\pm 0.99$&	$\le 0.61$\\
3c286	&  4&	0.6922&	-1.20&	$19.50	\pm 2.25$&	$5.27	\pm 1.47$\\
1107+0048& 5,6&0.7404&	-0.60&	$10.00	\pm 0.75$&	$10.72	\pm 1.24$\\
0454+039&  7&	0.8596&	-0.99&	$4.90	\pm 0.68$&	$2.14	\pm 0.32$\\
1727+5302& 5&	0.9449&	-0.52&	$14.50	\pm 4.35$&	$18.62	\pm 1.91$\\
0302-223&  7&	1.0093&	-0.54&	$2.29	\pm 0.59$&	$2.82    \pm 0.14$\\
1727+5302& 5&	1.0312&	-1.39&	$26.00	\pm 7.80$&	$4.47	\pm 0.46$\\
0515-44 &  3&	1.1500&	-0.95&	$2.82	\pm 1.00$&	$1.36	\pm 0.13$\\
0948+43	&  8& 1.2330& -0.95&	$31.62	\pm 7.35$&	$15.14	\pm 0.00$\\
0935+417&  9&	1.3726& -0.78&	$2.50	\pm 0.50$&	$1.76	\pm 0.41$\\
1354+258&  10&	1.4200&	-1.58&	$34.67	\pm 4.81$&	$3.89	\pm 0.97$\\
0933+733&  5&	1.4790&	-1.58&	$42.00	\pm 8.00$&	$4.68	\pm 1.47$\\
1104-180A& 11&	1.6614&	-0.99&	$7.08	\pm 0.16$&	$3.08	\pm 0.07$\\
1151+068&  12&	1.7736&	-1.53&	$20.00	\pm 5.00$&	$2.50	\pm 0.50$\\
1331+170&  13&	1.7764&	-1.26&	$15.00	\pm 1.42$&	$3.48	\pm 0.23$\\
2314-409&  14& 	1.8573&	-1.01&	$7.94	\pm 1.85$&	$3.31	\pm 0.77$\\
2230+025&  15&	1.8642&	-0.68&	$7.08	\pm 1.38$&	$6.31	\pm 0.41$\\
1210+1731& 13&	1.8920&	-0.86&	$3.98	\pm 0.92$&	$2.34	\pm 0.16$\\
2206-199&  13&	1.9205&	-0.37&	$4.50	\pm 0.74$&	$8.20	\pm 0.17$\\
1157+014&  16&	1.9440&	-1.34&	$63.10	\pm 14.66$&	$12.30	\pm 2.28$\\
0551-366&  17&	1.9620&	-0.11&	$3.16	\pm 0.59$&	$10.47	\pm 1.21$\\
0013-004&  18&	1.9733&	-0.72&	$6.70	\pm 0.70$&	$5.47    \pm 0.50$\\
1850+44&   19&	1.9900&	-0.68&	$25.12	\pm 5.84$&	$22.39	\pm 5.20$\\
1215+333& 13,15&1.9990&-1.25&$8.91	\pm 1.38$&	$2.14	\pm 0.24$\\
0010-002&  8&	2.0250&	-1.16&	$6.31	\pm 1.47$&	$1.86	\pm 0.28$\\
0458-020& 13,15&2.0395& -1.15&$44.67 \pm 9.32$&	$13.61	\pm 0.63$\\
2231-00	 & 15&	2.0660&	-0.73&	$3.63	\pm 0.84$&	$2.90	\pm 0.15$\\
0049-283&  20& 2.0713&	$\le -1.04$&	$3.10	\pm 0.40$&	$\le 1.22$\\
2206-199&  13& 2.0762&	$\le -1.86$&	$2.70	\pm 0.37$&	$\le 0.16$\\
2359-02	&  15&	2.0950&	-0.73&	$5.00	\pm 1.16$&	$3.94	\pm 0.26$\\
0149+33	&  15&	2.1400&	-1.64&	$3.16	\pm 0.73$&	$0.31	\pm 0.08$\\
0528-250&  21& 	2.1410&	-1.45&	$8.91	\pm 1.03$&	$1.35	\pm 0.45$\\
2359-02	&  15& 2.1540&	$\le -1.03$&	$2.00	\pm 0.46$&	$\le 0.80$\\
\tableline
\end{tabular}
\end{center}
\end{table}
\clearpage

\begin{table}
\centerline{\bf{TABLE 10. Summary of Zn Measurements in DLAs (cont'd)}}
\bigskip
\begin{center}
\begin{tabular}{|l|c|l|l|l|l|}
\tableline
QSO & Ref.$^{\dagger}$&  $z_{abs}$&[Zn/H]&$N_{\rm H I}/10^{20}$(cm$^{-2}$)& $N_{\rm Zn II}/10^{12}$(cm$^{-2}$)\\
\tableline
\tableline
1451+123 &22&	2.2550&	-1.08&	$2.00	\pm 0.70$&	$0.71	\pm 0.18$\\
2348-147& 15,20 &  2.2789&	$\le -1.28$&	$3.63	\pm 0.63$&	$\le 0.82$\\
0216+080& 23 &	2.2931&	$\le -0.27$&	$2.82	\pm 1.06$&	$\le 6.46$\\
PHL957	& 15 &	2.3091&	-1.54&	$25.10	\pm 2.90$&	$3.12	\pm 0.17$\\
2243-6031& 24&	2.3300&	-1.08&	$4.68	\pm 0.22$&	$1.66	\pm 0.12$\\
1232+0815& 25&	2.3376&	-0.84&	$8.00	\pm 1.00$&	$4.90	\pm 1.60$\\
0841+129& 15&	2.3745&	-1.47&	$8.91	\pm 1.80$&	$1.30	\pm 0.15$\\
0112+029& 20&	2.4227&	-1.13&	$9.00	\pm 2.00$&	$2.86	\pm 0.82$\\
2343+12	& 26&	2.4310&	-0.74&	$2.24	\pm 0.26$&	$1.73	\pm 0.05$\\
0836+113&13&    2.4651&	$\le -1.09$&	$3.80	\pm 0.88$&	$\le 1.32$\\
1223+178&13&	2.4658&	-1.58&	$31.62	\pm 7.35$&	$3.55	\pm 0.21$\\
CTQ247A	& 27 &	2.5505&	-1.32&	$13.49	\pm 3.13$&	$2.75	\pm 0.32$\\
1209+093& 28&	2.5840&	-1.05&	$25.12	\pm 5.84$&	$9.59	\pm 1.17$\\
CTQ247B	& 27&	2.5950&	-1.04&	$12.30	\pm 2.86$&	$4.79	\pm 0.22$\\
2348-011& 13&   2.6145&$\le -2.06$&	$19.95	\pm 4.63$&	$\le 0.74$\\
0913+072& 12&   2.6183&	$\le -1.09$&	$2.30	\pm 0.40$&	$\le 0.80$\\
1759+75	& 29&	2.6250&	-1.05&	$5.77	\pm 0.09$&	$2.20	\pm 0.42$\\
0812+32	& 28&	2.6260&	-0.95&	$22.40	\pm 5.20$&	$10.80	\pm 0.55$\\
Ctq460	& 28&   2.7770&$\le -1.27$&	$10.00	\pm 2.32$&	$\le 2.29$\\
0056+014& 12&	2.7771&	-1.20&	$13.00	\pm 2.00$&	$3.50	\pm 2.00$\\
1253-025& 28&	2.7830&	-1.71&	$70.79	\pm 33.77$&	$5.86	\pm 1.00$\\
1132+2243& 28&  2.7830&	$\le -1.64$&	$10.00	\pm 1.62$&	$\le 0.97$\\
1337+113&  28&  2.7950&	$\le -1.38$&	$8.91	\pm 2.07$&	$\le 1.59$\\
2138-444&   8&	2.8520&	-1.48&	$6.31	\pm 1.17$&	$0.89	\pm 0.21$\\
2342+34&   28& 	2.9080&	-1.23&	$12.59	\pm 0.29$&	$3.17	\pm 0.80$\\
1021+3001& 28&  2.9490&	$\le -1.14$&	$5.01	\pm 1.16$&	$\le 1.55$\\
0426-2202& 28&  2.9830&	$\le -1.97$&	$31.62	\pm 11.14$&	$\le 1.46$\\
0347-383& 30&   3.0250&-0.96&	$3.63   \pm 0.42$&	$1.70	\pm 0.20$\\
2334-09	& 28 &  3.0570&	$\le -0.91$&	$2.82	\pm 0.65$&	$\le 1.47$\\
0808+52	& 28 &  3.1130&	$\le -1.15$&	$4.47	\pm 0.72$&	$\le 1.36$\\
1535+2943& 31&  3.2020&$\le -0.75$&	$4.47	\pm 1.57$&	$\le 3.40$\\
2344+0342& 28&  3.2190&	$\le -1.71$&	$22.39	\pm 3.62$&	$\le 1.85$\\
2315+0921& 31&  3.2190&$\le -2.03$&	$22.39	\pm 7.89$&	$\le 0.89$\\
1506+522& 28&   3.2240&	$\le -1.20$&	$4.68    \pm 0.76$&	$\le 1.27$\\
1432+39	&  28&  3.2720&$\le -1.23$&	$17.78	\pm 4.13$&	$\le 4.42$\\
0957+33	&  28&  3.2800&$\le -0.95$&	$2.82	\pm 0.52$&	$\le 1.35$\\
2239-386&  12&  3.2810&$\le 	-1.01$&	$5.75	\pm 1.01$&	$\le 2.40$\\
2155+1358& 28&  3.3160&	$\le -1.13$&	$3.55	\pm 1.25$&	$\le 1.13$\\
\tableline
\end{tabular}
\end{center}
\end{table}
\clearpage

\begin{table}
\centerline{\bf{TABLE 10. Summary of Zn Measurements in DLAs (cont'd)}}
\bigskip
\begin{center}
\begin{tabular}{|l|c|l|l|l|l|}
\tableline
QSO & Ref.$^{\dagger}$&  $z_{abs}$&[Zn/H]&$N_{\rm H I}/10^{20}$(cm$^{-2}$)& $N_{\rm Zn II}/10^{12}$(cm$^{-2}$)\\
\tableline
\tableline

1715+3809& 31	& 3.3410&$\le -1.57$&	$11.20	\pm 3.47$&	$\le 1.30$\\
1117-1329& 32	&3.3504&-1.21&	$6.92	\pm 1.70$&	$1.81	\pm 0.12$\\
0000-263&  33	&3.3901&	-2.03&	$25.70	\pm 4.76$&	$1.02	\pm 0.12$\\
1802+5616& 31	& 3.3910&$\le -0.53$&	$2.00	\pm 0.46$&	$\le 2.54$\\
0007+2417& 31	&3.4960	&$\le -1.34$&	$12.60	\pm 2.92$&	$\le 2.45$\\
1802+5616& 31	&3.5540&$\le -0.50$&	$3.16    \pm 0.74$&	$\le 4.29$\\
1723+2243& 28	&3.6950&	-0.62&	$3.16	\pm 1.11$&	$3.22	\pm 1.07$\\
1248+31	 & 28   &3.6960&	$\le -0.65$&	$4.27	\pm 0.69$&	$\le 4.12$\\
1535+2943& 31	&3.7610&$\le 0.08$&	$2.51	\pm 0.89$&	$\le 12.90$\\
0133+0400& 28	&3.7740&$\le -0.08$&	$3.55	\pm 0.98$&	$\le 12.56$\\
0747+2739& 28	&3.9000&	$\le -0.73$&	$3.16	\pm 0.74$&	$\le 2.50$\\
\tableline
\end{tabular}
\end{center}
\hskip40pt {$\dagger$ References: 
 1. This Paper; 
 2. Boisse et al. (1998); 
 3. de la Varga et al. (2000);
 4. Meyer \& York (1992);
 5. Khare et al. (2004);  
 6. Rao, Turnshek, \& Nestor (2004); 
 7. Pettini et al. (2000);   
 8. Prochaska et al. (2003b); 
 9. Meyer et al. (1995);
 10. Pettini et al. (1999);  
 11. Lopez et al. (1999); 
 12. Pettini et al. (1997); 
 13. Prochaska et al. (2001b); 
 14. Ellison \& Lopez (2001); 
 15. Prochaska \& Wolfe (1999); 
 16. Petitjean, Srianand, \& Ledoux et al. (2000); 
 17. Ledoux, Srianand, \& Petitjean (2002); 
 18. Petitjean, Srianand, \& Ledoux et al. (2002); 
 19. Prochaska \& Wolfe (1998);  
 20. Pettini et al. (1994); 
 21. Centurion et al. (2003); 
 22. Dessauges-Zavadsky et al. (2003);
 23. Lu et al. (1996);  
 24. Lopez et al. (2002); 
 25. Ge, Bechtold, \& Kulkarni (2001); 
 26. Dessauges-Zavadsky et al. (2004); 
 27. Lopez \& Ellison (2003);
 28. Prochaska et al. (2003c); 
 29. Prochaska et al. (2002); 
 30. Ledoux, Petitjean, \& Srianand (2003); 
 31. Prochaska et al. (2003a); 
 32. Peroux et al. (2002);  
 33. Molaro et al. (2001).

} 
\end{table}
\clearpage

\begin{table}
\tabletypesize{\scriptsize}
\rotate
\centerline{\bf {TABLE 11}}
\centerline{\bf{Global Zn Metallicity vs. Redshift (Binned)}}
\vskip15pt
\begin{center}
\begin {tabular}{|c|c|c|c|c|c|c|c|}
\tableline
\small
$z$ range &  Med. &D,L$^{\dagger}$
&${\rm log}\,(\overline Z / Z_{\odot})$&${\rm log}\,(\overline Z / Z_{\odot})$&${{\rm log}\,(\overline Z / Z_{\odot})}$&${{\rm log}\,(\overline Z / Z_{\odot})}$
&${\rm log}\,(\overline Z / Z_{\odot})$\\
&$z$& &Max. Lim. & Min. Lim.& Min. Lim.&Mean of & Surv. Anal.\\
&&&Zn only&Zn only&Zn+others&Max, Min$^{\dagger \dagger}$&\\
\tableline
\tableline
&&&&&&&\\
0.09-1.37&0.74&10, 5&$-0.86$& $-1.01$ &$-0.93$&$-0.89$&$-0.94$ \\
&&&$\pm 0.11$&$\pm 0.14$&$\pm 0.10$&$\pm 0.12$&$\pm 0.16$\\
&&&&&&&\\

1.42-2.00&1.88&14, 0&$-1.07 $& $-1.07 $&$-1.07 $&$-1.07 $&$-1.07 $\\
&&&$\pm 0.17$&$\pm 0.17$&$\pm 0.17$& $\pm 0.17$&$\pm 0.16$\\
&&&&&&&\\

2.02-2.34&2.14&10, 5&$-1.09 $& $-1.19 $&$-1.14 $&$-1.12 $&$-1.16 $\\
&&&$\pm 0.13$&$\pm 0.10$&$\pm 0.10$&$\pm 0.12$&$\pm 0.23$\\
&&&&&&&\\

2.37-2.78&2.59&10, 4&$-1.20 $& $-1.25 $&$-1.22 $&$-1.21 $&$-1.21$\\
&&&$\pm 0.08$&$\pm 0.10$&$\pm 0.08$&$\pm 0.08$&$\pm 0.14$\\
&&&&&&&\\

2.78-3.27&3.02&4, 11&$-1.48 $& $-1.92 $&$-1.69 $&$-1.59 $&$-1.65 $\\
&&&$ \pm 0.14$& $ \pm 0.23$&$ \pm 0.14$&$ \pm 0.20$&$ \pm 0.19$\\
&&&&&&&\\

3.28-3.90&3.44&3, 11&$-0.86$& $-1.80 $&$-1.43 $&$-1.14 $&$-1.30 $\\
&&&$\pm 0.25$& $\pm 0.51$&$\pm 0.22$&$\pm 0.46$&$\pm 0.15$\\

\tableline
\end{tabular}
\end{center}
\hskip40pt {$\dagger$ D: Number of Zn Detections, L: Number of Zn Limits. 
$\dagger \dagger$: The minimum limit value used while calculating this mean  
is the one with information from other elements used in cases of Zn limits.}
\end{table}
\clearpage

\begin{table}
\centerline{\bf {TABLE 12}}
\centerline{\bf{Exponential Fits to Global Metallicity-Redshift Relation}}
\vskip15pt
\begin{center}
\begin {tabular}{|l|l|l|c|c|}
\tableline
Redshift range&Binning&Treatment of Limits &${\rm log}\,(\overline Z_{0} / Z_{\odot})
$&$B$ \\
\tableline
\tableline
0.09-3.90&Binned & Maximum limits & $-0.74 \pm 0.15$ & $-0.18 \pm 0.06$\\
0.09-3.90&Binned & Minimum limits, Zn only & $-0.75 \pm 0.18$ & $-0.22 \pm 0.08$\\
0.09-3.90&Binned & Minimum limits, Zn + others$^{\dagger}$&$-0.71 \pm 0.13$&$-0.23 \pm 0.06$\\
0.09-3.90&Binned & Mean of Max., Min. Limits$^{\dagger}$&$-0.72 \pm 0.16$&$-0.20 \pm 0.07$\\
0.09-3.90&Binned & K-M Survival Analysis & $-0.79 \pm 0.18$ & $-0.18 \pm 0.07$\\ 
0.09-3.28&Binned & Maximum limits & $-0.66 \pm 0.15$ & $-0.23 \pm 0.07$\\
0.09-3.28&Binned & Minimum limits, Zn only&$-0.77 \pm 0.18$&$-0.22 \pm 0.08$\\
0.09-3.28&Binned & Minimum limits, Zn + others$^{\dagger}$ & $-0.69 \pm 0.14$ & $-0.24 \pm 0.06$\\
0.09-3.90&Unbinned &Maximum limits & $-0.86 \pm 0.15$  & $-0.11 \pm 0.08$\\
0.09-3.90&Unbinned &Minimum limits, Zn only & $-0.95 \pm 0.08$&$-0.14 \pm 0.05$\\
0.09-3.90&Unbinned &Minimum limits, Zn + others $^{\dagger}$&  $-0.86 \pm 0.08$ & $-0.15 \pm 0.05$\\
\tableline
\end{tabular}
\end{center}
\hskip40pt {$\dagger$: Here, the minimum limit 
calculation refers to the one with information from other elements used in cases 
of Zn limits.}
\end{table}
\clearpage

\end{document}